\documentclass[twocolumn]{aastex631}

\usepackage{amsmath}
\usepackage{bm}
\usepackage{verbatim}
\usepackage{booktabs}
\usepackage{hyperref}
\usepackage{multirow}
\usepackage{threeparttable}

\newcommand{\vesc}{v_{\rm esc}}
\newcommand{\vcirc}{v_{\rm circ}}
\newcommand{\vmin}{v_{\rm min}}

\newcommand{\kms}{\rm{km\ s^{-1}}}

\newcommand{\msun}{M_{\odot}}
\newcommand{\rsun}{r_{\odot}}
\defcitealias{Leonard1990}{LT90}

\shorttitle{}
\shortauthors{}

\begin{document}


\title{Impact of Gaia-Sausage-Enceladus Debris on Milky Way Escape-velocity Estimates in TNG50}

\correspondingauthor{Haining Li}
\email{lhn@nao.cas.cn}

\author[0009-0003-4116-6824]{Yin Wu}
\affiliation{National Astronomical Observatories, Chinese Academy of Sciences, Beijing 100101, People’s Republic of China}
\affiliation{School of Astronomy and Space Science, University of Chinese Academy of Sciences, Beijing 100049, People’s Republic of China}

\author[0000-0002-0389-9264]{Haining Li}
\affiliation{National Astronomical Observatories, Chinese Academy of Sciences, Beijing 100101, People’s Republic of China}

\author[0000-0001-8382-6323]{Shi Shao}
\affiliation{National Astronomical Observatories, Chinese Academy of Sciences, Beijing 100101, People’s Republic of China}
\affiliation{School of Astronomy and Space Science, University of Chinese Academy of Sciences, Beijing 100049, People’s Republic of China}

\author[0000-0002-0642-5689]{Xiang-Xiang Xue}
\affiliation{National Astronomical Observatories, Chinese Academy of Sciences, Beijing 100101, People’s Republic of China}
\affiliation{Institute for Frontiers in Astronomy and Astrophysics, Beijing Normal University, Beijing 102206, People’s Republic of China}

\author[0000-0001-7080-0618]{Lan Zhang}
\affiliation{National Astronomical Observatories, Chinese Academy of Sciences, Beijing 100101, People’s Republic of China}

\author[0000-0002-8980-945X]{Gang Zhao} 
\affiliation{National Astronomical Observatories, Chinese Academy of Sciences, Beijing 100101, People’s Republic of China}
\affiliation{School of Astronomy and Space Science, University of Chinese Academy of Sciences, Beijing 100049, People’s Republic of China}

\begin{abstract}

The escape velocity is a key probe of the Milky Way (MW) potential, but standard estimates can be biased by unrelaxed substructures.
Using MW analogs from the TNG50 cosmological simulation, we investigate the impact of the dominant, highly radial Gaia-Sausage-Enceladus (GSE) debris on escape-velocity measurements.
We introduce an adaptive selection strategy that balances statistical reliability with sample size to identify the power-law decline regime of the local halo velocity distribution.
Applying this strategy to 16 MW analogs with prominent GSE-like structures and continuous high-velocity tails, we find that retaining GSE-like stars yields a systematic underestimate, with a recovery fraction of $97.5^{+2.1}_{-2.9}$\%, while removing GSE-like stars alone increases the recovery fraction to $99.4^{+2.1}_{-2.5}$\%. 
The $\sim 2\%$ underestimation of the escape velocity corresponds to an $\sim 6\%$ underestimation of the total mass.
This improvement persists in LAMOST-like magnitude and footprint selection.
GSE-like debris is therefore a key contributor to the systematic underestimation of escape-velocity measurements, with direct implications for the inferred MW mass.

\end{abstract}


\section{Introduction} 
\label{sec:intro}

In the $\Lambda$CDM paradigm, galaxies grow through hierarchical accretion and mergers of lower-mass systems \citep{Press1974,White1978}.
The Milky Way (MW) is no exception to this scenario, as its stellar halo preserves tidal debris from multiple accretion events, including disrupted dwarf galaxies \citep[e.g.,][]{Helmi1999,Bullock2005,Belokurov2006}.
Therefore, the kinematics of these remnants provides a powerful probe of MW dynamics \citep[for recent reviews, see][]{Helmi2020,Deason2024,Hunt2025}.
This context motivates a reexamination of methods used to infer MW properties, specifically the escape velocity.

The escape velocity, defined as the maximum velocity for a star bound to the gravitational field at a given location, is a measure of the potential depth of the MW.
The seminal work of \citet[][hereafter \citetalias{Leonard1990}]{Leonard1990} proposed a standard method to measure the escape velocity by modeling the velocity distribution of high-velocity stars as a truncated power law.
Combined with a mass model of the MW, this formalism has been extensively applied to estimate the escape velocity and thereby constrain the dark matter (DM) mass and the overall mass distribution \citep[e.g.,][]{Smith2007,Piffl2014,Williams2017,Monari2018,Deason2019,Koppelman2021,Necib2022a,Prudil2022,Roche2024,Wu2025}.
However, the theoretical assumptions underlying the \citetalias{Leonard1990} method have been increasingly challenged.

One of the most critical issues is that the \citetalias{Leonard1990} method relies on a smooth, well-mixed stellar distribution, which may be violated by the complex, unrelaxed Galactic stellar halo \citep{Grand2019}.
Thanks to the massive astrometric data from the Gaia mission \citep{Lindegren2016,Gaia2023},
combined with other large spectroscopic surveys such as 
RAVE \citep{Steinmetz2006},
the Sloan Digital Sky Survey \citep[SDSS,][]{York2000}, 
LAMOST \citep{Cui2012,Zhao2012,Yan2022}, 
APOGEE \citep{Majewski2017}, 
GALAH \citep{Silva2015},
numerous dynamical substructures have
been identified \citep[e.g.,][]{Belokurov2018,Haywood2018,Helmi2018,Koppelman2019,Matsuno2019,Myeong2019,Naidu2020,Yuan2020,Malhan2022,Malhan2024}.
These substructures strongly overlap in the local phase space \citep[e.g.,][]{Naidu2020,Dodd2023,Horta2023,Zhang2024,Xie2025}, introducing kinematic complexity.
Moreover, systems with substructures exhibit high-velocity tails that differ from those of a smooth, fully relaxed halo \citep[e.g.,][]{Grand2019,Koppelman2021,Necib2022b,Roche2024}.

Among the various substructures, Gaia-Sausage-Enceladus \citep[GSE;][]{Belokurov2018,Haywood2018,Helmi2020} stands out as a dominant component of ex situ stars in the MW \citep[e.g.,][]{Nelson2019a,Naidu2020,Naidu2021}.
Such debris, with highly radial orbits and high anisotropy, can populate the high-velocity tail \citep{Deason2019,Necib2019}.
Since escape-velocity estimates are sensitive to the shape of the high-speed tail, it is essential to validate the \citetalias{Leonard1990} method in systems with GSE structures.

Motivated by recent findings of GSE-like debris in TNG50 MW analogs \citep{Folsom2025},
we investigate how GSE-like structures bias escape-velocity estimates using MW analogs in \citet{Pillepich2024}.
In Section~\ref{sec:data}, to construct our mock solar neighborhood samples, we isolate GSE-like stars using purely kinematic criteria \citep[e.g.,][]{Naidu2020,Lane2022,Carrillo2024} and incorporate LAMOST magnitude and footprint selection \citep{Cui2012,Zhao2012,Yan2022} to represent realistic survey data.
We introduce an adaptive selection of the high-velocity tail and the modeling process of $\vesc$ in Section~\ref{sec:method}.
By comparing the modeled $\vesc$ with the true $\vesc$ of the simulation,
we rigorously evaluate the extent to which removing these GSE-like structures can recover the true $\vesc$ and total mass, as described in Section~\ref{sec:result} and Section~\ref{sec:discussion}.
Finally, we summarize our findings in Section~\ref{sec:conclusion}.

\section{Data} \label{sec:data}


\subsection{TNG50 Simulations and MW Analogs}

For our study, we used the high-resolution TNG50 simulation \citep{Pillepich2019,Nelson2019b} of the IllustrisTNG project 
\citep{Marinacci2018,Naiman2018,Pillepich2018b,Springel2018,Nelson2019a}.
The simulation is based on the \citet{Planck2016} cosmology, with a Hubble constant of $H_0=100h$ km s$^{-1}$ Mpc$^{-1}=67.74$ km s$^{-1}$ Mpc$^{-1}$, and evolved using the moving mesh code \textsc{arepo} \citep{Springel2010}.
The initial conditions are generated in the early Universe (redshift $\approx127$) and evolved to the present day (redshift $= 0$), following a comprehensive model of galaxy formation
\citep{Weinberger2017,Pillepich2018a}.
The TNG50 simulation operates in a periodic cubic volume with a side length of 51.7 cMpc, containing 2160$^3$ DM particles and an equal initial number of gas cells.
The mass resolutions are $4.5\times10^5\;\msun$ and $8.5\times10^4\;\msun$ for DM and baryonic particles \citep{Pillepich2024}, respectively, which are crucial for properly resolving low-mass merger remnants with stellar masses of order $10^7$--$10^9\;\msun$ and provide new opportunities to constrain the formation and assembly histories of MW-mass galaxies.

We made use of the catalogs proposed by \citet{Pillepich2024}, 
who identify 198 MW and M31 analogs at redshift $=0$ according to stellar mass ($M_{\star}$), disk morphology, and environment.
Since we focus only on MW analogs, an additional stellar mass cut ($M_{\star}<10^{10.9}\;\msun$) was applied, resulting in 159 MW analogs.
For these analogs, we extracted essential data for all particles and cells, such as coordinates, velocities, masses, and stellar photometry \citep{Nelson2019a}.
The coordinates and velocities were converted to the coordinate system of the main galaxy, with more details in \citet{Pillepich2024}.
We excluded particles belonging to satellite galaxies, which have been studied in \citet{Engler2021} and \citet{Engler2023}, and focused only on the remaining stars in the main galaxies for the subsequent dynamical and stellar property calculations.

To separate stellar populations based on their origins, we utilized \textsc{Sublink} merger trees \citep{Rodriguez-Gomez2015} to trace particles from redshift $=0$ back to their formation time.
Following the definitions described in \citet{Rodriguez-Gomez2016}, stars formed within the main progenitor branch are classified as in situ, which can be effectively distinguished from ex situ stars that formed in satellite galaxies and were subsequently accreted into the host.

\subsection{Properties of MW Analogs with GSE-like Features} \label{subsec:properties}

\begin{figure*}
   \centering
   \includegraphics[width=0.90\textwidth]{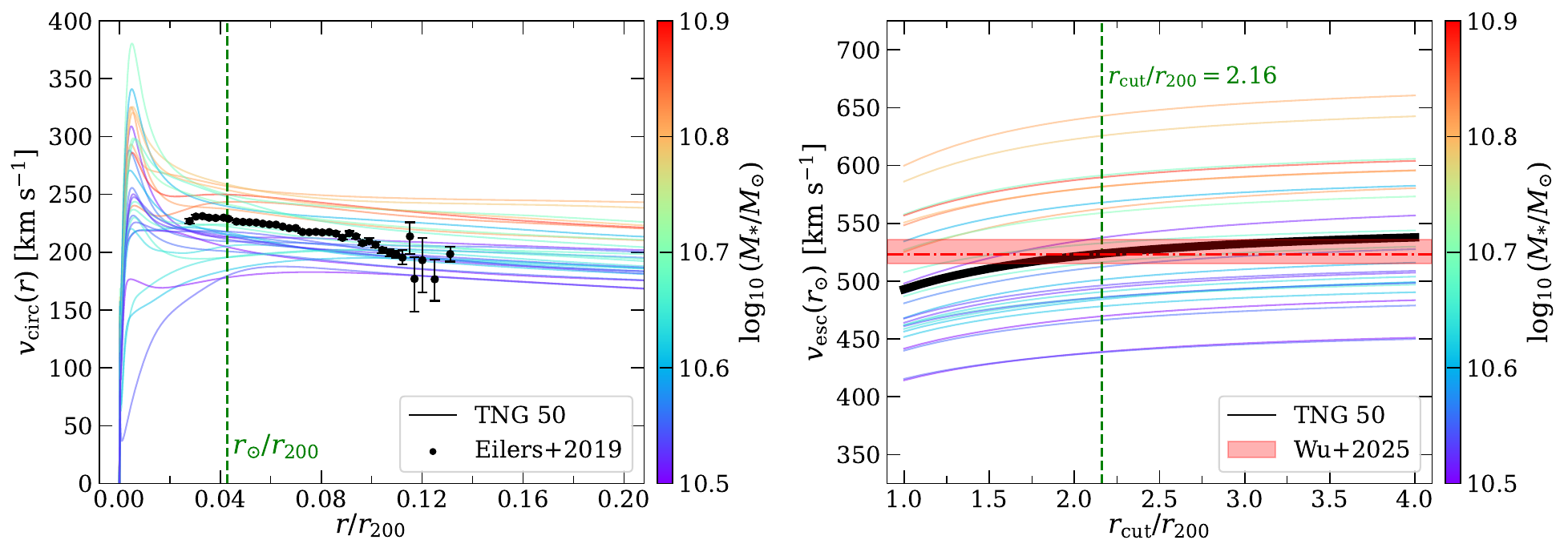}
   \caption{Left: circular velocity curves for 28 MW analogs with prominent GSE-like structures. We add the derived circular velocity curve (black dots) and mark the solar radius (green dashed line) in \citet{Eilers2019}. Right: local escape velocity, $\vesc(\rsun)$, as a function of $r_{\rm cut}$ for these 28 MW analogs, with the median value represented by a black solid line. We also plot the most recent local escape-velocity estimates in \citet{Wu2025} as the red area. Curves in both panels are color coded by stellar mass, spanning a range of $\sim 10^{10.5}$--$10^{10.9}\;\msun$. All radii are normalized by the virial radii of these analogs.}
   \label{fig:tng_mw_curves}
   \vspace{1.0em}
\end{figure*}

\begin{table*}
\begin{center}
\caption{Selection Criteria for MW Analogs with Prominent GSE-like Structures.}\label{tab:samples}
\begin{tabular}{lc}
  \hline\hline
   Selection Criteria & Number of Galaxies \\
  \hline
  MW/M31 analogs from \citet{Pillepich2024} & 198 \\
  $M_{\star}<10^{10.9}\;\msun$ & 159 \\
  Mass fraction of the GSE-like stars to ex situ stars $\gtrsim 50$\% & 28 \\
   Reliable escape-velocity modeling & 16 \\
  \hline
\end{tabular}
\end{center}
\end{table*}

The methodology for deriving dynamics from TNG50 simulations is more straightforward than in observations,
as the mass distribution of each type of particle is directly accessible. 
In this study, we adopted two right-handed coordinate systems: a cylindrical system ($R$, $\phi$, $z$) and a spherical system ($r$, $\theta$, $\phi$), 
both centered on the main galaxy.
The gravitational potential $\Phi(r)$ of the MW analogs was constructed with the \textsc{agama} software package \citep{Vasiliev2019} under the assumption of approximate spherical symmetry, 
using the positions and masses of all DM, gas and stellar particles.
A detailed discussion on the impact of adopting different symmetry assumptions is provided in Appendix~\ref{appendix:A}.
We then computed the orbital parameters, including orbital energy ($E$), azimuthal action ($J_{\phi}=-Rv_\phi$), circularity ($\eta={J_{\phi}}/{|J_{\rm circ}(E)|}$), orbital apocenter ($r_{\rm apo}$) and pericenter ($r_{\rm peri}$), and eccentricity ($e=\dfrac{r_{\rm apo}-r_{\rm peri}}{r_{\rm apo}+r_{\rm peri}}$).
Here, $J_{\rm circ}(E)$ is the angular momentum of a circular orbit with the same energy $E$.
We define negative $v_\phi$ as corresponding to disk rotation; thus, positive $J_{\phi}$ corresponds to prograde orbits.

While a variety of chemical and dynamical criteria have been employed in the literature to identify GSE stars \citep[e.g.,][]{Feuillet2020,Naidu2020,Lane2022,Horta2023,Lane2023,Carrillo2024,Zhang2024}, selections relying on energy or angular momentum are inherently sensitive to the varying potential depths of host galaxies.
In this study, we select GSE-like stars by combining intrinsic ex situ classification with present-day kinematics,
i.e., ex situ stars with high eccentricity ($e>0.7$) and low azimuthal velocity ($|v_\phi|<100$ km s$^{-1}$), 
in order to select a more complete GSE-like sample \citep[e.g.,][]{Lane2022,Carrillo2024}.
Furthermore, to ensure prominent GSE-like structures, 
we selected analogs in which the mass fraction of GSE-like stars to ex situ stars is $\gtrsim 50$\% \citep[e.g.,][]{Naidu2020,Naidu2021}.
The ex situ classification is an intrinsic simulation label and is not directly observable. 
In MW observations, accreted populations can instead be identified using chemical information, such as the [Mg/Mn] versus [Al/Fe] plane \citep[e.g.,][]{Horta2023}, combined with kinematics \citep[e.g.,][]{Matsuno2021,Zhang2024,Xie2025}.
However, applying the same criteria to TNG50 is nontrivial because the predicted abundance distributions depend on the chemical-evolution model, the adopted selection boundaries, and observational uncertainties. 
We therefore use intrinsic ex situ labels to define a clean, idealized baseline for isolating the dynamical effect of accreted debris.

Furthermore, we impose no explicit constraint on the infall times of the GSE-like progenitors. 
Accretion times cannot be measured directly from observations, 
and systems selected by their present-day GSE-like kinematics span lookback times of $\sim 3$--$11$ Gyr \citep[see also][]{Folsom2025}.
This range extends to more recent epochs than the commonly inferred $8$--$11$ Gyr lookback-time interval for the MW's GSE accretion event \citep[e.g.,][]{Helmi2018,Naidu2021,Donlon2024}.
Nevertheless, our subsequent goodness-of-fit criterion naturally disfavors systems containing strongly unrelaxed debris from very recent mergers.

We note that our selected GSE-like stars may originate from multiple merger events, consistent with recent studies suggesting that GSE stars may comprise debris from more than one progenitor \citep[e.g.,][]{Orkney2023,Folsom2025}.
Regardless of their accretion histories, these stars occupy phase-space regions similar to observed GSE populations, making them suitable proxies for GSE debris when evaluating their effects on the escape velocity.
Our final sample contains 28 MW analogs with prominent GSE-like debris, 19 of which overlap with the sample of \citet{Folsom2025}.
Table~\ref{tab:samples} summarizes the selection criteria and the number of the resulting galaxies.
In practice, not all MW analogs have continuous high-velocity tails that allow statistically reliable $\vesc$ fits.
We therefore restrict our analysis to the valid subset of 16 galaxies, as detailed in Section~\ref{subsec:multiple} and Appendices~\ref{appendix:B} and \ref{appendix:C}.

As noted in previous work, the TNG50 MW analogs typically exhibit more extended disks 
\citep[scale lengths $R_{\rm d}=4.4_{-1.8}^{+3.2}$ kpc; e.g.,][]{Sotillo-Ramos2022,Pillepich2024} than the MW, which has $R_{\rm d,\;MW}=1.7$--$2.9$ kpc \citep[e.g.,][]{Hammer2007,Juric2008,Bovy2013,Bland-Hawthorn2016}.
Local kinematic and mass distributions are poorly reproduced at a fixed physical radius of 8.2 kpc \citep{Bland-Hawthorn2016}.
Consequently, we compared the circular velocity curves calculated by $v_{\rm circ}(r)=\sqrt{GM(<r)/r}$ with observations to determine the physical solar neighborhood.

The left panel of Figure~\ref{fig:tng_mw_curves} presents the circular velocity curves for 28 MW analogs with prominent GSE-like structures in our study.
In most galaxies, the circular velocity curves become flat at the solar radius, $r_{\odot}/r_{200}\approx 0.043$ in \citet{Eilers2019}.\footnote{We define $r_{200}$ as the radius within which the average galaxy density is 200 times the critical density and $M_{200}$ as the mass enclosed in $r_{200}$.}
Therefore, we adopted the regime corresponding to $|r-8.2\;\rm{kpc}|<1.0\;\rm{kpc}$ in the MW, i.e., $0.038<r/r_{200}<0.048$, as the physically defined solar neighborhood.\footnote{Unless otherwise noted, we use this definition throughout our study.}
We did not impose an additional cut on $|z|$. 
As a robustness check, we repeated the analysis for the representative well-mixed system (\textsc{Subfind} ID: 505100) using the cuts of $|z|<2$ kpc and $|z|<5$ kpc. 
The inferred escape velocities are consistent with the result obtained without a vertical cut, although the vertical cuts lead to broader constraints due to the reduced sample sizes.

The local escape velocity at the solar volume is widely defined as 
\begin{equation}
  v_{\rm esc}(r)=\sqrt{2|\Phi(r)-\Phi(r_{\rm cut})|},
  \label{eq:vesc_sph}
\end{equation}
where $r_{\rm cut}$ is an arbitrary limiting radius for realistic, nonisolated galaxies like the MW.
Unlike an isolated system, the gravitational potential in cosmological simulations does not vanish at infinity, implying that the escape velocity should be defined relative to a finite reference radius.
Previous studies have generally taken $r_{\rm cut}$ as $2r_{200}$
\citep[e.g.,][]{Deason2019,Necib2022a,Roche2024,Wu2025} or 
$3r_{340}$ \citep[e.g.,][]{Piffl2014,Monari2018,Koppelman2021}.

To assess the sensitivity of the local escape velocity to the reference radius,
we calculated $\vesc(r_{\odot})$ over a range of $r_{\rm cut}$ values and compared with the most recent measurements of $\vesc(r_{\odot})=523.74^{+12.83}_{-13.47}\;\kms$ \citep{Wu2025}, as illustrated in the right panel of Figure~\ref{fig:tng_mw_curves}.
The comparison indicates that the median simulated escape velocity is consistent with the observed value at $r_{\rm cut}\approx2.16\times r_{200}$.
This radius is consistent with the region where the simulated $v_{\rm esc}(r_\odot)$ profile begins to flatten.
Beyond this radius, the local escape velocity changes little with increasing $r_{\rm cut}$ ($\lesssim10\;\mathrm{km\,s^{-1}}$ within the range of $2$--$3r_{200}$), 
indicating that matter at larger radii contributes only weakly to the local gravitational potential.
This convergence provides independent support for our adopted $r_{\rm cut}$, allowing it to reasonably serve as a physical boundary of the simulated galaxies.
Different choices of $r_{\rm cut}$ would shift the absolute recovery fractions but would not affect the relative comparison between the ``with GSE'' and ``no GSE'' scenarios.

\begin{figure*}
   \centering
   \includegraphics[width=0.99\textwidth]{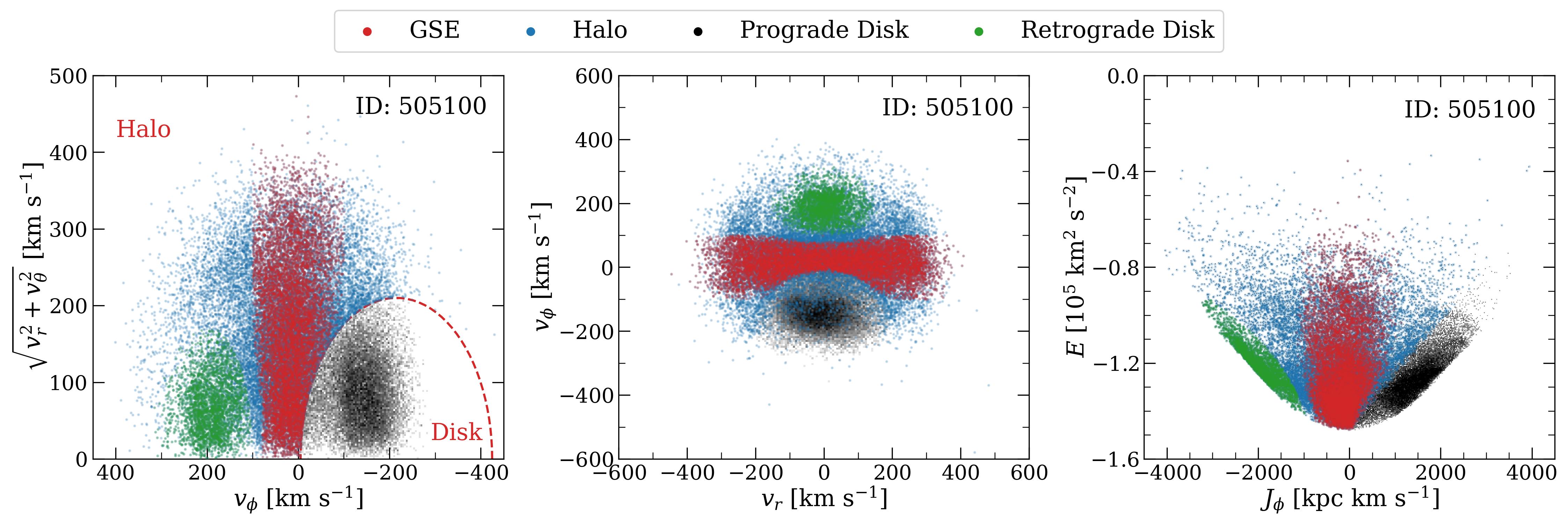}
   \caption{The distributions of local stars for an example MW analog (\textsc{Subfind} ID: 505100). The left panel displays the Toomre diagram, with the red dashed line representing the division between halo and disk populations. The distributions of these stars in the $v_{\phi}$--$v_{r}$ and $E$--$J_{\phi}$ planes are shown in the middle and right panels, respectively. The distinct stellar components are color coded: GSE-like stars (red), halo stars (blue), the primary prograde disk (black), and a retrograde disk component (green).}
   \label{fig:dynamics}
   \vspace{1.5em}
\end{figure*}

\subsection{Samples for Local Escape-velocity Estimates} \label{subsec:samples}

We then constructed samples for local escape velocity estimates.
Since the \citetalias{Leonard1990} method focuses on halo stars, we identified the stellar halo of the MW analogs using the Toomre diagram \citep[see, e.g.,][]{Helmi2018,Koppelman2019,Dodd2023}, defined as:
\begin{equation}
|\boldsymbol{v}-\boldsymbol{v}_{\rm LSR}|>210\;{\rm km\;s^{-1}}, 
\end{equation}
where $\boldsymbol{v}_{\rm LSR}$ is the velocity of the local standard of rest (LSR).
Since our sample of MW analogs spans a very narrow circular velocity range ($\vcirc\sim 200$--$250\;\kms$, see Table~\ref{tab:mw_analogs}), adopting a fixed velocity threshold yields results nearly identical to a scaled cut while maintaining direct consistency with MW observational works.
Figure~\ref{fig:dynamics} displays the distribution of local stars in kinematic space for a typical MW analog (\textsc{Subfind} ID: 505100), with $\vcirc(\rsun)=215.05$ $\kms$. 

The left panel shows that the halo and disk populations are distinctly separated by the Toomre diagram in velocity space. 
The middle and right panels present the selected GSE-like stars in the $v_\phi$--$v_r$ and $E$--$J_{\phi}$ planes, which confirm that the GSE-like particles cluster around $J_{\phi}\sim 0$ while spanning a wide range of orbital energies \citep[see also][]{Naidu2020,Dodd2023,Horta2023,Zhang2024,Xie2025}.
In addition to the dominant halo and GSE-like structures, we can also identify a less prominent retrograde disk component with $\eta<-0.7$.
We note that this component has a negligible impact on the validity of our subsequent escape velocity measurements for this galaxy.

In order to mimic observational effects, we also took LAMOST as an example and generated subsets with its selection strategy.
We converted the simulated intrinsic photometry to apparent magnitudes in the SDSS $r$ band \citep{Nelson2019a}.\footnote{We did not include any dust extinction for simplicity.}
Applying the standard selection criteria of the LAMOST survey, decl.$>-10^{\circ}$ and $9.0<r_{\rm mag}<17.5$ \citep{Cui2012,Zhao2012,Yan2022}, we defined our ``LAMOST sample'', which allows for a direct comparison with the complete ``All sample''.
Here, the apparent magnitude of each star particle is used as a practical proxy for the detectability of the stellar population it represents,
since our intention is to assess the statistical impact caused by the LAMOST-like observational selection, rather than constructing a fully realistic synthetic survey via full sampling.

\section{Methodology} \label{sec:method}

In this section, we present our framework for modeling $\vesc$ and investigating the impact of GSE-like debris.
First, we introduce a statistical method to evaluate when this model is valid.
Then, we isolate the impact of GSE by comparing fits from stellar subsets with and without GSE-like stars.
We also consider observational selection effects to provide practical insights into observations.

\subsection{Escape-velocity Modeling}
\label{subsec:vesc_modeling}

As introduced in Section~\ref{sec:intro}, the standard \citetalias{Leonard1990} approach assumes velocity distributions following
 \begin{equation}
  p(v| v_{\rm esc}, k)\propto (v_{\rm esc} - v)^k, 
  \label{eq:lt90}
\end{equation}
for $v_{\rm min} \leq v< v_{\rm esc}$, 
where $\vesc$ is the escape velocity, $k$ is the exponent, and $\vmin$ is the low-velocity threshold.
Since the fraction of unbound stars in the TNG50 simulated galaxies is negligible, we did not include the outlier component adopted in recent observational works \citep[e.g.,][]{Williams2017,Necib2022a,Roche2024,Wu2025}.
Therefore, the total likelihood function is given as:
\begin{equation}
   {\mathcal{L}}_{\rm total}=\prod_{i}^{N_{\rm stars}}p_i(v_i|v_{\rm esc}, k).
  \label{eq:L}
\end{equation}

We utilized a Bayesian framework to infer the free parameters $\vesc$ and $k$.
Our baseline setup assumed a uniform prior on $k$, $\mathcal{U}(0.1<k<10)$, and a widely used prior $\propto 1/\vesc$ over $\vmin<\vesc<1000$ km s$^{-1}$, similar to previous works \citep{Koppelman2021,Necib2022a,Prudil2022}.
To estimate the posterior distributions of $\vesc$ and $k$, the Markov Chain Monte Carlo (MCMC) sampling was performed in \textsc{emcee} \citep{Foreman2013}, using 50 walkers, 1000 steps as a burn in, and additional 2000 steps for further sampling.

However, one of the most critical steps in this Bayesian inference is to determine an appropriate $\vmin$.
While the large volume of the simulation data allows for well-converged posterior distributions \citep{Piffl2014}, 
the validity of the fitted parameters depends on whether the adopted $\vmin$ correctly identifies the regime where the truncated power law holds.
Unlike observational works that often choose a fixed $v_{\min}$ \citep[e.g.,][]{Monari2018,Deason2019,Wu2025} for the solar neighborhood, such an assumption is not suitable for our study.
The velocity distribution in simulated galaxies varies from galaxy to galaxy and may differ from that of the real MW.
The reliability of the truncated power law for TNG50 MW analogs therefore warrants further verification.

To address this, we implemented an adaptive $\vmin$-selection strategy.
We scanned candidate $\vmin$ values over a 100 km s$^{-1}$ interval with a step size of 2 km s$^{-1}$, setting the upper limit between 320 and 380 $\kms$ to retain a statistically large number of high-velocity GSE-like stars.
For each trial $\vmin$, we modeled the velocity distribution of local halo stars with $v>\vmin$ to derive the parameters $\vesc$ and $k$.
We then evaluated the goodness of fit by calculating the Kolmogorov–Smirnov (KS) $p$-value, comparing the modeled distribution with that of the simulated data.
As $v_{\min}$ increases and the velocity distribution enters the power-law decline regime, the $p$-value is expected to undergo a sharp initial increase.
Our fiducial $v_{\min}$ was then chosen as the velocity at which this initial rise ends, which can be identified as the first prominent peak with the $p$-value $>0.05$, as detailed in Section~\ref{subsec:single}.
This strategy retains the largest possible sample while ensuring that the high-velocity tail has already entered a stable power-law decline, balancing a valid fit with sufficient statistics.
Since observed high-velocity tails consistently exhibit a clear power-law decline \citep[e.g.,][]{Monari2018,Koppelman2021,Prudil2022,Necib2022b,Roche2024,Wu2025}, systems that produce valid fits are considered closer to the MW observations.

During the search process, the fitted parameters $\vesc$ and $k$ typically show a plateau near the fiducial $\vmin$, with variations of $\Delta \vesc<5$ $\kms$ and/or $\Delta k<0.1$ in most well-modeled galaxies.
This behavior implies that our method can indeed identify a stable power-law decline of high-velocity tails.
For systems with sparse high-velocity tails, however, posterior distributions can become strongly degenerate at high $\vmin$.
In these cases, we obtained the unified inference using the posterior of $k$ at the lower $\vmin$ as the prior at the higher $\vmin$.

\subsection{Comparative Analysis}
\label{subsec:comparison}

To systematically evaluate the impact of GSE-like structures and observational limitations, we considered four distinct scenarios: retaining (``with GSE'') versus removing (``no GSE'') the GSE-like stars, applied across both the complete ``All sample'' and the observationally motivated ``LAMOST sample''.
Applying our pipeline to these scenarios, we obtained the posterior distribution of $\vesc$ to directly compare with the true local $\vesc$ values calculated in Section~\ref{subsec:properties}.

Furthermore, we propagated these $\vesc$ estimates into a mass modeling to assess the impact of GSE on mass inference.
We modeled the total potential by combining a fixed baryonic component with a Navarro-Frenk-White \citep[NFW;][]{Navarro1996} DM halo.
The baryonic potential of each simulated galaxy was constructed directly via multipole expansion in the \textsc{agama} software package \citep{Vasiliev2019}, 
using the positions and masses of all gas and stellar particles.
MCMC sampling was then performed using \textsc{emcee} \citep{Foreman2013} to constrain the DM virial mass ($M_{\rm DM,200}$) and the concentration ($c_{\rm DM,200}$), using the full posterior distribution of $\vesc$.
During the MCMC sampling, we fixed the contribution of the baryonic component and varied only the parameters of the NFW DM halo.
The total virial mass was then calculated by summing the baryonic mass and the DM virial mass for each MW analog.
Given the significant deviation of the inner DM halo from an NFW profile in the TNG cosmological simulation \citep[e.g.,][]{Lovell2018}, we added the circular velocity at $0.1r_{200}$ instead of the local circular velocity as an additional constraint in the likelihood evaluation.

\section{Results} \label{sec:result}

\subsection{Validation on a Single Galaxy} 
\label{subsec:single}

\begin{figure*}
   \centering
   \includegraphics[width=0.95\textwidth]{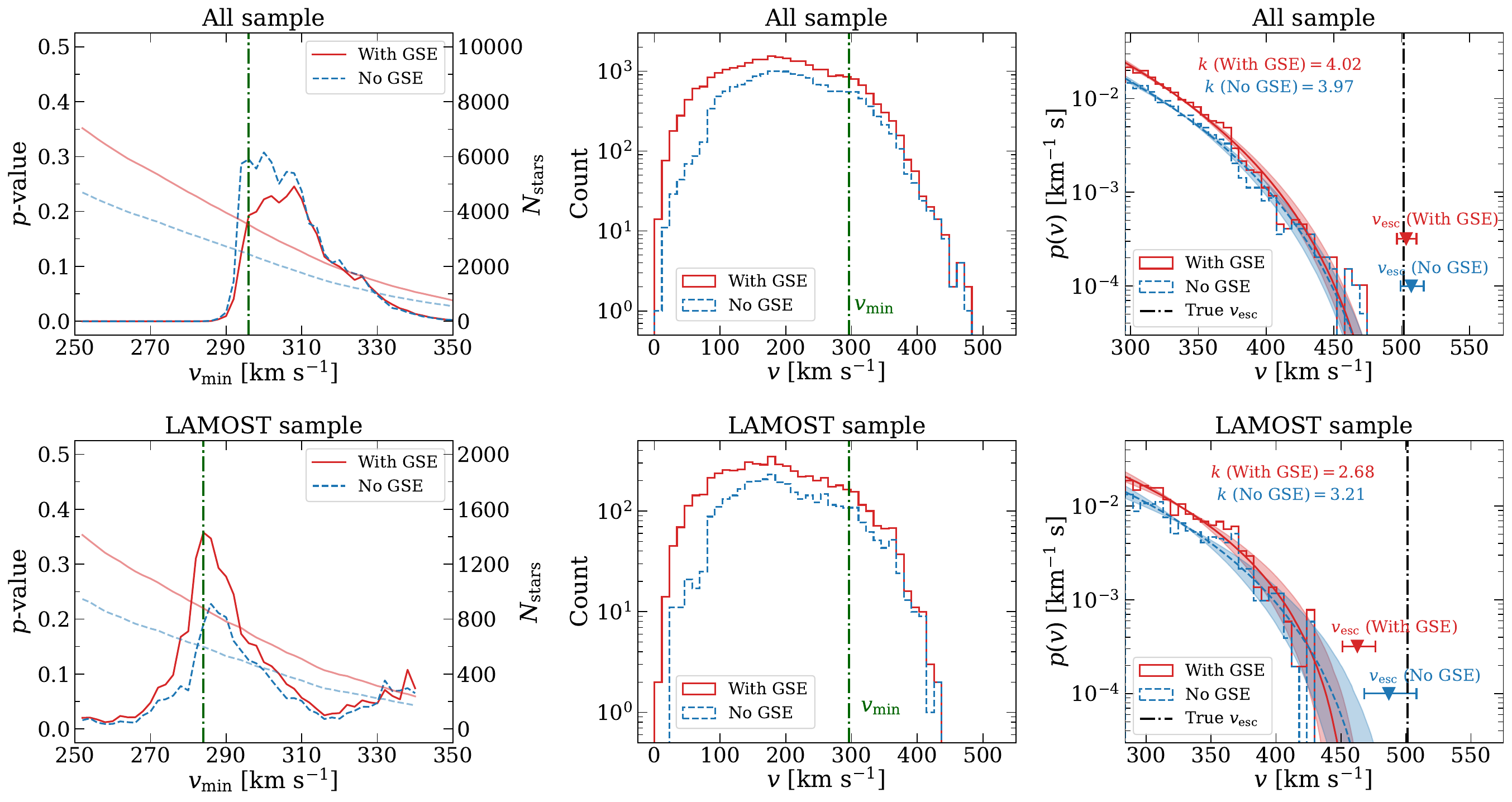}
   \caption{The adaptive selection of the fiducial $\vmin$ and velocity distribution modeling for a typical MW analog (\textsc{Subfind} ID: 505100). Top and bottom rows display results for the ``All sample'' and ``LAMOST sample'', respectively. Across all panels, red and blue colors denote the ``with GSE'' and ``no GSE'' scenarios. The fiducial $\vmin$ is marked as the green line. Left: the $p$-value and sample size ($N_{\rm stars}$) as a function of $\vmin$. Middle: the velocity distributions over the entire velocity range. Right: the comparison between the best fits and velocity distributions of high-velocity tails, with the true $\vesc$ (black lines) and the fitted $\vesc$ (scatter point with error bars) marked.}
   \label{fig:test_505100}
   \vspace*{1.2em}
\end{figure*}

To validate our adaptive $\vmin$-selection strategy, we present the results for a typical MW analog with a relatively smooth velocity distribution (\textsc{Subfind} ID: 505100) in Figure~\ref{fig:test_505100}.
The left panels show the results of our adaptive $\vmin$ selection.
For both the ``All sample'' and ``LAMOST sample'', the $p$-value exhibits an initial rise followed by a drop as $\vmin$ increases.
Since this trend remains highly consistent regardless of whether the sample contains GSE-like stars, we adopted the same fiducial $\vmin$ for the ``with GSE'' and ``no GSE'' scenarios.

As described in Section~\ref{subsec:vesc_modeling}, the initial rise indicates that the velocity distribution has entered the region governed by a power-law decline.
The subsequent drop indicates that when $\vmin$ is too high, the statistical reliability is compromised due to the significant decrease in the sample size.
The extreme high-velocity tail is particularly sensitive to discrete, unrelaxed substructures. 
Even in this well-mixed system, the distribution fluctuates more severely at extremely high velocities,
which reduces the goodness of fit to a smooth power-law model.

In the overall velocity distributions (middle panels of Figure~\ref{fig:test_505100}), our fiducial $\vmin$ corresponds to the onset of a stable power-law decline, i.e., 
beyond this threshold, the exponent $k$ remains relatively constant.
The final modeling results shown in the right panels reproduce the decreasing trend of the high-velocity tails well, 
indicating that our $\vmin$-selection strategy identifies an appropriate $\vmin$ consistent with the definition in \citetalias{Leonard1990}.

\subsection{Results of Multiple Milky Way Analogs}
\label{subsec:multiple}
 
We then expanded our adaptive $\vmin$-selection pipeline to the full suite of 28 MW analogs (see Appendices~\ref{appendix:B} and \ref{appendix:C} for details).
Among these systems, 16 galaxies with relatively continuous and smooth high-velocity tails yield reliable escape velocity fits.
By contrast, our pipeline naturally rejects the remaining 12 systems with high-velocity tails that have obvious clumpy features and thereby deviate from the single power-law model, as shown in Appendix~\ref{appendix:B}.
For such systems with clumpy velocity distributions, an alternative approach is to introduce multiple power-law components to account for distinct substructures \citep[e.g.,][]{Necib2022b}.
However, since our primary objective is to evaluate the reliability of the standard \citetalias{Leonard1990} method, which assumes a smooth, well-mixed stellar halo, we do not introduce additional power-law components, allowing our pipeline to naturally filter out unrelaxed systems.
Taken together, these results indicate that our pipeline provides a robust diagnostic for identifying systems with dynamically smooth stellar halos.

In Appendix~\ref{appendix:C}, even among the 16 galaxies with reliable escape-velocity modeling, the accuracy of the \citetalias{Leonard1990} method depends sensitively on the smoothness of the high-velocity tails \citep[see also][]{Grand2019}. 
In some galaxies (e.g., \textsc{Subfind} IDs: 498522, 538905, and 540082), the true $\vesc$ lies in the 1$\sigma$ confidence intervals of the fitted $\vesc$ for both the ``with GSE'' and ``no GSE'' scenarios.
This behavior suggests that the velocity distribution of GSE-like components is consistent with that of the underlying smooth halos in these systems, i.e., GSE-like structures may be well-mixed in the host halos.

However, in other systems whose velocity distributions exhibit multiple small bumps introduced by substructures (e.g., \textsc{Subfind} IDs: 492876, 504559, and 530330), GSE-like structures substantially reshape the high-velocity tails, producing more extended tails and hence lower $k$ values \citep[see also][]{Deason2019,Necib2022a,Roche2024}.
Removing the GSE-like stars leads to the escape-velocity estimates which are more consistent with the true values. 
A detailed discussion of the $\vesc$-modeling bias caused by GSE-like structures is provided in Section~\ref{sec:discussion}.

Comparing the modeling results for the ``All sample'' and the ``LAMOST sample'', we note that observational selection effects may randomly change the smoothness of the high-velocity tails (e.g., \textsc{Subfind} IDs: 454172 and 530330).
However, the systematic bias caused by GSE-like structures and the improvement from removing GSE-like stars still persist.

\section{Discussion} \label{sec:discussion}

\begin{figure*}[htbp]
   \centering
   \includegraphics[width=0.99\textwidth]{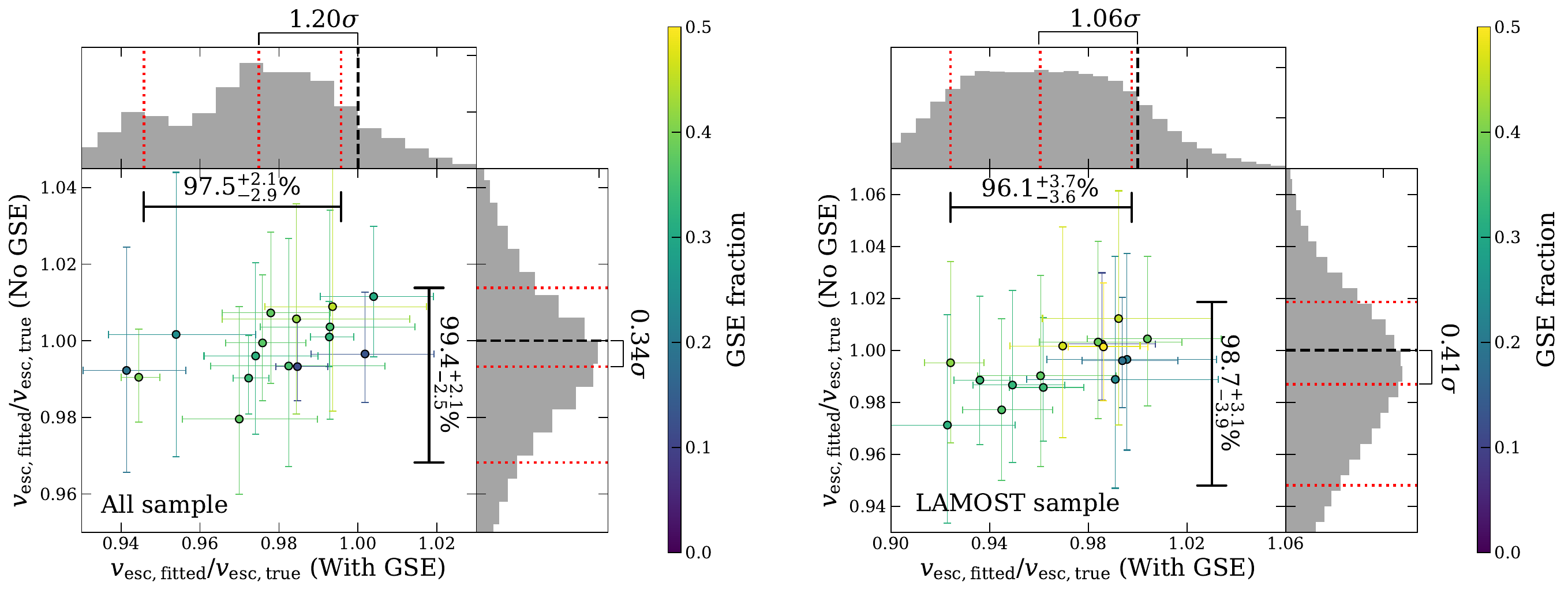}
   \caption{The comparison of the recovery fraction, $v_{\rm esc,\;fitted}/v_{\rm esc,\;true}$ for the ``All sample'' (left panels) and the ``LAMOST sample'' (right panels). The marginalized distributions show the joint posterior of the ``with GSE'' (\textit{x}-axis) and ``no GSE'' (\textit{y}-axis) scenarios. In these histograms, the black dashed lines indicate the ground truth (1.0), while the red dotted lines mark the median and the 68\% confidence intervals. Each scatter point represents an individual MW analog, with error bars denoting the 68\% confidence intervals and color indicating the fraction of GSE-like stars within the high-velocity tail.}
   \label{fig:fits_stats}
   \vspace*{1.5em}
\end{figure*}

To quantify the systematic bias induced by GSE-like structures, we construct joint posterior distributions for 16 final galaxies and compare the fitting results between the ``with GSE'' and the ``no GSE'' scenarios. 
Figure~\ref{fig:fits_stats} shows the comparison of the recovery fraction, given by $v_{\rm esc,\;fitted}/v_{\rm esc,\;true}$, with the 1$\sigma$ confidence intervals defined by the 16th and 84th percentiles.
Among these galaxies, the proportion of GSE-like stars within high-velocity tails spans a broad range, from $\sim$10\% to 50\%.
We note that the recovery fraction is insensitive to the proportion of GSE-like stars,
confirming that the systematic bias is driven by differences in the intrinsic dynamical properties rather than by the relative fraction of the unrelaxed substructures.

As shown in the left panels of Figure~\ref{fig:fits_stats}, the marginalized distribution for the ``with GSE'' scenario of the ``All sample'' displays a broad, noticeably left-skewed profile, yielding a systematic underestimate with a recovery fraction of $97.5^{+2.1}_{-2.9}$\% and a deviation of 1.2$\sigma$ from the true values.
This underestimate can be physically attributed to the dynamical properties of the substructures.
Stars from a single merger event, such as the GSE-like merger, possess distinct binding energies and phase-space distributions, failing to populate high-velocity tails efficiently and continuously \citep[e.g.,][]{Grand2019,Necib2019,Necib2022b}.
Such strongly unrelaxed debris that produces localized bumps or discontinuities is therefore generally rejected by our goodness-of-fit criterion.
However, early merger debris may become phase mixed sufficiently to produce a smooth velocity distribution while still retaining dynamical properties distinct from those of the underlying stellar halo.
In this case, the debris can contribute continuously over a broad velocity range without causing the single power-law model to be rejected.
The fitting algorithm accommodates these components by adopting a flatter decline (lower $k$), which inherently biases the best-fit $v_{\rm esc}$ towards lower values.
This explains the observed skewed distribution in the ``with GSE'' scenario even after systems with the most prominent bumpy substructures have been excluded.

Similar underestimates have also been reported in the Auriga simulations \citep[see, e.g.,][]{Grand2019,Koppelman2021}. 
Our inferred offset is smaller ($\sim$2\%) compared to $\sim$7\% in \citet{Grand2019} or $\sim$10\% in \citet{Koppelman2021}.
In addition to differences in simulation resolution and in the exact truncation location of high-velocity tails, this reduction likely reflects the influence of our pipeline, which selects only systems with robust $\vesc$ fits.

Crucially, as discussed in Section~\ref{subsec:multiple}, removing GSE-like structures effectively mitigates this bias.
For the ``no GSE'' scenario of the ``All sample'', the right marginalized distribution in the left panel of Figure~\ref{fig:fits_stats} shows a more symmetric shape.
The recovery fraction of the true $\vesc$ increases to $99.4^{+2.1}_{-2.5}$\%, with a reduced deviation of 0.34$\sigma$.
This confirms that GSE-like structures can bias inferred escape velocities under the well-mixed assumption of the \citetalias{Leonard1990} method and serve as an important and likely leading source of systematic bias among the systems considered here.
Indeed, our findings are consistent with the established physical picture that dominant GSE structures with highly radial orbits \citep[e.g.,][]{Belokurov2018,Haywood2018,Helmi2018,Naidu2020,Naidu2021} can populate high-velocity tails \citep{Deason2019,Necib2019}.

The corresponding recovery fraction for the ``LAMOST sample'' is presented in the right panel of Figure~\ref{fig:fits_stats}.
Observational selection effects manifest as broadening and flattening the overall posterior distribution, thereby erasing some intrinsic statistical features.
However, despite this flattening effect, the improvement in the removal of GSE-like structures remains significant. 
The recovery fraction increases from $96.1^{+3.7}_{-3.6}$\% to $98.7^{+3.1}_{-3.9}$\%, with a deviation decreasing from 1.06$\sigma$ to 0.41$\sigma$.
Therefore, our pipeline, which combines the adaptive $\vmin$-selection strategy with the removal of GSE-like stars, remains effective under this simplified observational selection.
It effectively mitigates intrinsic physical biases and remains insensitive to observational selection effects, providing a reliable framework for accurately estimating $\vesc$.

\begin{figure}[htbp]
   \centering
   \includegraphics[width=0.43\textwidth]{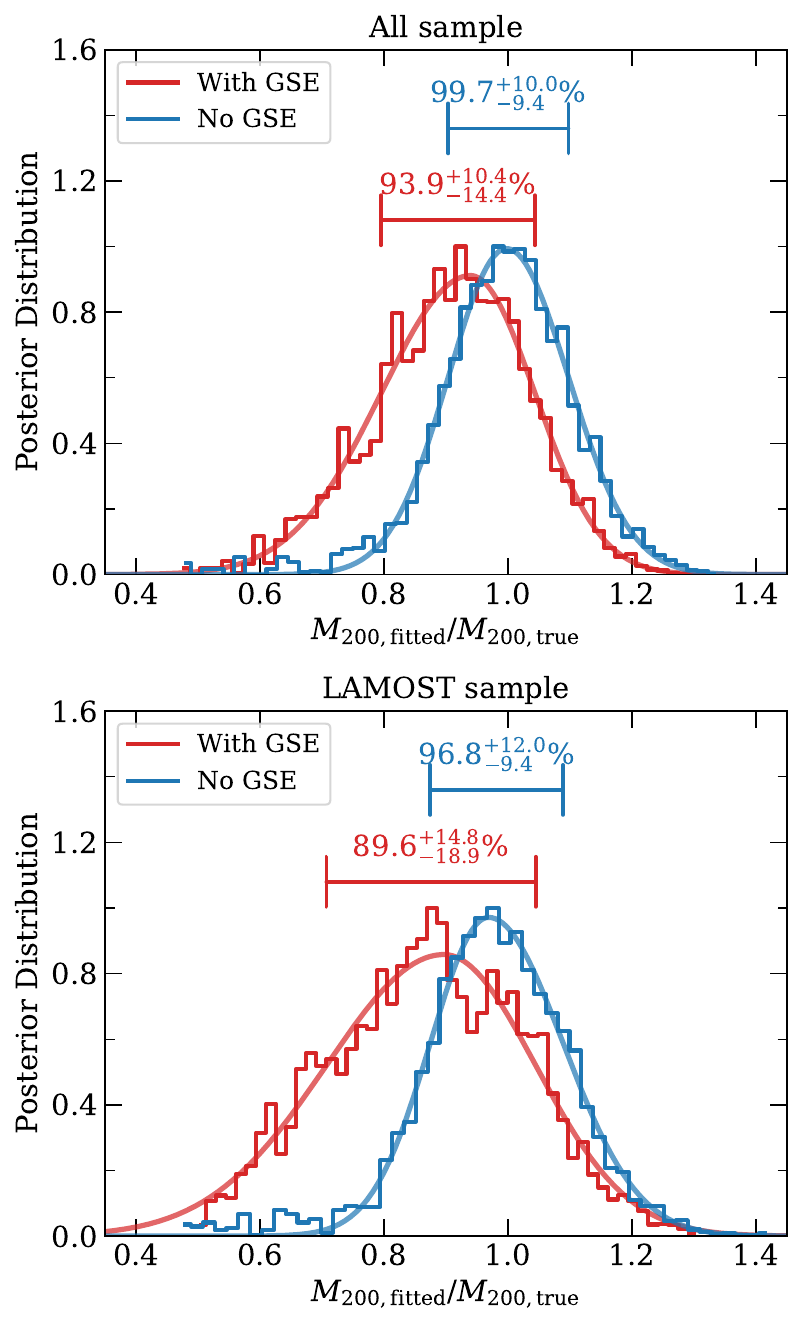}
   \caption{Posterior distributions of the mass recovery fraction, $M_{\rm 200, fitted}/M_{\rm 200, true}$. The top and bottom panels show the results from the ``All sample'' and the ``LAMOST sample'', respectively. Red and blue colors denote the ``with GSE'' and ``no GSE'' scenarios. The smooth curves show the best-fit asymmetric Gaussian models, from which the median and the 1$\sigma$ credible intervals are derived.}
   \label{fig:fits_mass}
   \vspace*{1.0em}
\end{figure}

Furthermore, we demonstrate the posterior distributions of the mass recovery fraction, $M_{\rm 200,\;fitted}/M_{\rm 200,\;true}$, in Figure~\ref{fig:fits_mass}.
To better demonstrate the posterior distribution of the mass, we also modeled it using an asymmetric Gaussian model and derived the best-fit median and 1$\sigma$ credible intervals.
Comparing the results from the ``with GSE'' and the ``no GSE'' scenarios, the mass estimates show substantial overlap within the 1$\sigma$ confidence intervals, both consistent with the values of the true total mass.
Nevertheless, the results of the ``no GSE'' scenarios can better recover the true total mass.
The $\sim 2\%$ underestimation of $\vesc$ results in an $\sim 6\%$ underestimation of the total mass, indicating the necessity to consider the impact of GSE more carefully when estimating the total mass of the MW by the escape-velocity method.

\section{Conclusions} \label{sec:conclusion}

In this work, we used TNG50 MW analogs to quantify how GSE-like debris biases local escape-velocity measurements based on the \citetalias{Leonard1990} method.
From 159 MW analogs in \citet{Pillepich2024}, we selected 28 systems with prominent GSE-like structures, and 16 of them yielded statistically valid fits with continuous high-velocity tails.
Our main conclusions are as follows.

\begin{enumerate}
  \item We develop an adaptive $\vmin$-selection strategy that combines statistical reliability with sample size. 
  This approach robustly identifies the velocity range in which the truncated power-law model is valid, and provides a reliable diagnostic for dynamically smooth stellar halos. 
  This step is essential for identifying the subset of halos for which LT90-based inference is meaningful.

  \item The accuracy of the inferred escape velocity is intrinsically related to the smoothness of the stellar halo's velocity distribution. 
  Complex assembly and evolution histories of MW analogs result in unrelaxed substructures and introduce clumpy features in the stellar velocity distribution, thereby invalidating the single power-law model.

  \item Although multiple substructures may have an impact on high-velocity tails, GSE, as the most kinematically extreme and dominant substructure, is an important source of systematic underestimation in escape-velocity measurements for MW-like systems. 
  Specifically, removing GSE-like stars alone significantly increases the recovery fraction of the true escape velocity ($v_{\rm esc,\;fitted}/v_{\rm esc,\;true}$) from $97.5^{+2.1}_{-2.9}$\% to $99.4^{+2.1}_{-2.5}$\% for the complete ``All sample''.

  \item The same trend persists after applying observational selection effects, such as the LAMOST-like magnitude and footprint selection. 
  Despite broader posteriors from smaller sample sizes, removing GSE-like stars still improves the recovery fraction from $96.1^{+3.7}_{-3.6}$\% to $98.7^{+3.1}_{-3.9}$\% for the ``LAMOST sample''. Since the results of the ``All sample'' have higher recovery fractions and narrower 1$\sigma$ confidence intervals, it is necessary to obtain more accurate escape velocity measurements using all-sky coverage data.

  \item We also investigated the impact of GSE on the mass estimates. In this study, an $\sim 2\%$ underestimate of the escape velocity leads to an $\sim 6\%$ underestimate of the total mass for MW-like systems. Therefore, accounting for GSE substructures is essential for an accurate recovery of the Galactic potential and total mass.

\end{enumerate}

\section*{Acknowledgements}

We thank the anonymous referee for very efficient and helpful suggestions to improve the draft.
This work is supported by 
the National Natural Science Foundation of People’s Republic of China (NSFC) grant No. 12588202, 12633005, and 12273053,
the National Key $R\&D$ Program of People’s Republic of China Nos. 2024YFA1611902, 2024YFA1611903 and 2023YFE0107800, and 
the CAS Project for Young Scientists in Basic Research (Nos. YSBR-092 and YSBR-062), 
the Strategic Priority Research Program of Chinese Academy of Sciences grant No. XDB1160100, 
the science research grants from the People’s Republic of China Manned Space Project with Nos. CMS-CSST-2021-B03 and CMS-CSST-2025-A11, 
and National Astronomical Observatories Chinese Academy of Sciences Nos. E5ZB0301, E4ZB0301, E6510301 and E6510302.

The TNG50 simulation was run with compute time granted by the Gauss Center for Supercomputing (GCS) under Large-Scale Projects GCS-DWAR on the GCS share of the supercomputer Hazel Hen at the High Performance Computing Center Stuttgart (HLRS).

\software{\textsc{agama} \citep{Vasiliev2019}, astropy \citep{Astropy2013,Astropy2018,Astropy2022}, emcee \citep{Foreman2013}, matplotlib \citep{Hunter2007}, numpy \citep{numpy}, scipy \citep{scipy}, TNG access scripts \citep{Nelson2021}.}

\appendix

\section{Impact of Different Potential Approximations} \label{appendix:A}

\setcounter{table}{0}
\counterwithin*{table}{part}
\renewcommand{\thetable}{A\arabic{table}}

\begin{deluxetable*}{cccccccccc}[htbp]
    \tablecaption{Properties of 28 Milky Way analogs with Prominent GSE-like Structures}
    \label{tab:mw_analogs}
    \tablehead{
         \colhead{Index} & \colhead{\textsc{Subfind} ID} & \colhead{$M_{\star}$} & \colhead{$r_{200}$} & \colhead{$M_{200}$} & \colhead{$\vcirc(\rsun)$} & \colhead{$\vesc^{\rm sph}(\rsun)$} & \colhead{$\vesc^{\rm axi}(\rsun)$} & \colhead{$f_{\rm local,\;GSE}$} & \colhead{Status/Label} \\
          & & [$10^{10}\;\msun$] & [kpc] & [$10^{12}\;\msun$] & [$\kms$] & [$\kms$] & [$\kms$] &  &
    }
    \startdata
    1 & 446665 & 3.24 & 253.11 & 1.74 & 212.96 & 537.56 & 540.16 & 0.61 & Valid \\
    2 & 454172 & 3.63 & 248.18 & 1.62 & 215.45 & 512.41 & 514.88 & 0.49 & Valid \\
    3 & 465255 & 6.17 & 285.78 & 2.51 & 258.43 & 625.70 & 628.79 & 0.72 & Valid \\
    4 & 492876 & 6.92 & 263.10 & 1.95 & 243.53 & 581.70 & 586.41 &0.66 & Valid \\
    5 & 496186 & 4.90 & 262.50 & 1.95 & 249.81 & 591.46 & 594.05 &0.66 & Valid \\
    6 & 498522 & 7.59 & 259.81 & 1.86 & 249.87 & 589.79 & 595.49 &0.42 & Valid \\
    7 & 504559 & 3.98 & 257.82 & 1.82 & 239.57 & 567.93 & 570.55 &0.62 & Valid \\
    8 & 505100 & 3.98 & 249.72 & 1.66 & 215.05 & 501.28 & 503.16 & 0.63 & Valid \\
    9 & 515296 & 5.01 & 245.63 & 1.58 & 241.35 & 558.94 & 561.17 &0.35 & Valid \\
    10 & 528322 & 3.16 & 207.45 & 0.95 & 204.42 & 469.67 & 470.54 &0.64 & Valid \\
    11 & 530330 & 4.68 & 234.77 & 1.38 & 222.20 & 524.12 & 528.24 &0.70 & Valid \\
    12 & 538905 & 4.27 & 218.43 & 1.12 & 205.39 & 490.68 & 496.33 &0.52 & Valid \\
    13 & 540082 & 4.07 & 207.68 & 0.95 & 217.06 & 484.24 & 486.77 &0.64 & Valid \\
    14 & 540920 & 3.39 & 222.47 & 1.17 & 211.07 & 496.24 & 503.99 &0.80 & Valid \\
    15 & 541218 & 3.31 & 223.16 & 1.17 & 213.44 & 493.88 & 496.57 &0.80 & Valid \\
    16 & 552414 & 4.57 & 203.72 & 0.91 & 195.54 & 485.98 & 490.28 &0.62 & Valid \\
    17 & 422754 & 6.46 & 305.66 & 3.02 & 257.03 & 642.79 & 645.41 &0.63 & A \\
    18 & 440407 & 6.61 & 258.00 & 1.82 & 237.28 & 562.59 & 564.95 &0.64 & A \\
    19 & 511920 & 6.31 & 245.69 & 1.58 & 253.67 & 581.63 & 584.24 &0.74 & A \\
    20 & 512425 & 3.55 & 240.28 & 1.48 & 209.41 & 524.01 & 527.63 &0.56 & A \\
    21 & 534628 & 4.68 & 223.36 & 1.20 & 223.92 & 516.35 & 521.10 &0.55 & B \\
    22 & 537941 & 4.37 & 207.18 & 0.95 & 195.73 & 485.63 & 492.34 &0.61 & C \\
    23 & 543114 & 4.68 & 216.65 & 1.10 & 247.63 & 532.83 & 536.26 &0.38 & B \\
    24 & 543729 & 4.07 & 208.18 & 0.95 & 184.64 & 478.24 & 485.16 &0.61 & A \\
    25 & 550149 & 3.55 & 204.46 & 0.91 & 200.73 & 466.43 & 470.35 &0.64 & A \\
    26 & 552581 & 3.47 & 202.85 & 0.89 & 217.89 & 486.09 & 488.22 &0.67 & B \\
    27 & 565251 & 2.69 & 188.43 & 0.71 & 178.28 & 438.87 & 442.52 &0.29 & A \\
    28 & 571633 & 3.47 & 182.58 & 0.65 & 179.88 & 438.47 & 445.29 &0.34 & C \\
    \enddata

   \tablecomments{Columns from left to right are: (1) index, (2) \textsc{Subfind} ID from the TNG50 simulation at redshift $=0$, (3) stellar mass, (4) virial radius, (5) virial mass, (6) local circular velocity, (7) and (8) local escape velocity for spherical and axisymmetric approximations, respectively, (9) mass fraction of GSE-like stars to ex situ halo stars within the solar volume, and (10) fit status or the main reason for invalid fits using the \citetalias{Leonard1990} method. Systems with reliable fits are marked as ``Valid''. For systems with invalid fits, the labels are defined as follows: (A) clumpy high-velocity tails caused by recent massive mergers; (B) prominent contamination from retrograde disk-like structures; and (C) highly disk-dominated systems (disk fraction $>$ 70\% ) with insufficient high-velocity halo stars.}
\end{deluxetable*}

Table~\ref{tab:mw_analogs} summarizes the fundamental properties of the 28 MW analogs with prominent GSE-like structures.
To check on the impact of different potential approximations, we construct the gravitational potential of the MW analogs under the axisymmetric approximation.
The procedure is implemented using multipole expansion for the positions and masses of all DM, gas and stellar particles, with the \textsc{agama} software package \citep{Vasiliev2019}.
Then we calculate the local escape velocity using the modification of Equation~(\ref{eq:vesc_sph}), given by:
\begin{equation}
  v_{\rm esc}(R,\;z)=\sqrt{2|\Phi(R,\;z)-\Phi(r_{\rm cut})|},
  \label{eq:vesc_modified}
\end{equation}
where $\Phi(r_{\rm cut})$ represents the average potential of the spherical shell at $r_{\rm cut}$.
The local position is set to $(R,\;z)=(r_{\rm \odot},\;0)$, and the results are also listed in Table~\ref{tab:mw_analogs}.

We note that the typical error introduced by our spherical potential approximation is $<1\%$, and the spherical approximation tends to slightly underestimate the local escape velocity.
Furthermore, when re-evaluating the recovery of the true escape velocity under the axisymmetric approximation, the recovery fraction of the escape velocity in the all-sky scenario increases from $96.8\%$ to $98.7\%$ after removing the GSE-like structures. 
Similarly, when considering LAMOST-like selection effects, the recovery fraction improves from $95.4\%$ to $98.0\%$. 
Since the improvement of $\sim 2\%-3\%$ is consistent with that under the spherical approximation, we can conclude that different potential approximations have a negligible impact on our final results.

\section{MW analogs with Invalid Fits} \label{appendix:B}

\setcounter{table}{0}
\counterwithin*{table}{part}
\renewcommand{\thetable}{B\arabic{table}}

\setcounter{figure}{0}
\counterwithin*{figure}{part}
\renewcommand{\thefigure}{B\arabic{figure}}

\begin{figure*}[t]
   \centering
   \includegraphics[width=0.9\textwidth]{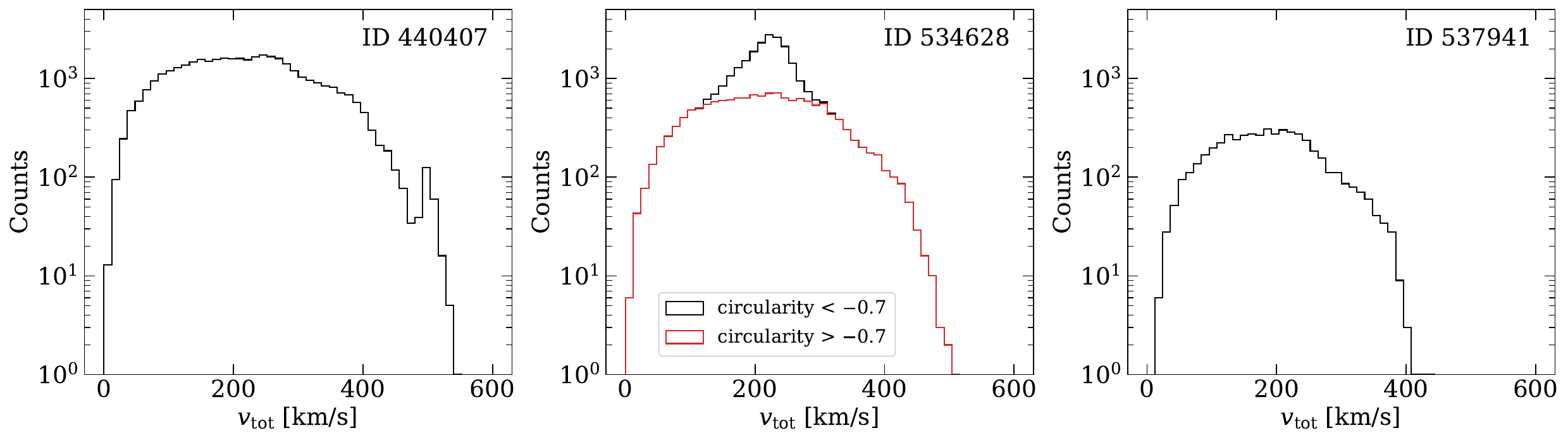}
   \caption{Local stellar velocity distributions over the entire range for three typical examples of MW analogs with invalid fits, \textsc{Subfind} ID: 440407, 534628, and 537941.}
   \label{fig:bad_examples}
\end{figure*}

Table~\ref{tab:mw_analogs} also summarizes the fitting status of the 28 MW analogs.
Among these, 16 galaxies exhibit continuous and relatively smooth high-velocity tails, yielding reliable $\vesc$ fits under the \citetalias{Leonard1990} formalism, i.e., the $p$-value $>0.05$ at the fiducial $v_{\rm min}$.
The remaining 12 systems fail to obtain valid fitting results.
Figure~\ref{fig:bad_examples} presents representative examples of these invalid cases, which can be attributed primarily to three distinct kinematic origins.

First, the dominant reason is unrelaxed kinematic substructures, as shown in the left panel of Figure~\ref{fig:bad_examples}.
Some galaxies have experienced relatively recent massive mergers ($\lesssim 4$ Gyr);\footnote{The merger time is defined as the time when the progenitor last crossed the $r_{200}$.} thus, the accreted components are not fully mixed in the stellar halo.
Unrelaxed accreted components may produce clumpy, discontinuous high-velocity tails that violate the smooth-halo assumption of the \citetalias{Leonard1990} method.
Notably, several systems with recent massive mergers, e.g., \textsc{Subfind} ID 530330, still exhibit relatively smooth halo velocity distributions that are sufficient to yield valid fits.

Second, the middle panel of Figure~\ref{fig:bad_examples} exhibits the velocity distribution of a MW analog with a prominent retrograde disk-like structure, which is kinematically defined as circularity $<-0.7$.
Since disk and halo populations possess fundamentally different velocity distributions, prominent retrograde disk-like structures significantly distort the morphology of the high-velocity tail. 
Consequently, future precise measurements of the MW's escape velocity may require careful consideration of these retrograde structures.

Finally, a few MW analogs are highly disk-dominated, in which over $70\%$ of the stars belong to the disk population.
These systems lack a prominent stellar halo; thus, we are unable to extract a statistically significant sample of local high-velocity halo stars.
Furthermore, as shown in the right panel of Figure~\ref{fig:bad_examples}, the fastest halo stars do not reach the $\sim 0.9\times v_{\rm esc}(r_\odot)$ threshold, which also limits the escape velocity modeling.

\section{Escape Velocity Fitting Results} \label{appendix:C}

\setcounter{figure}{0}
\counterwithin*{figure}{part}
\renewcommand{\thefigure}{C\arabic{figure}}

\begin{figure*}[t]
   \centering
   \includegraphics[width=0.65\textwidth]{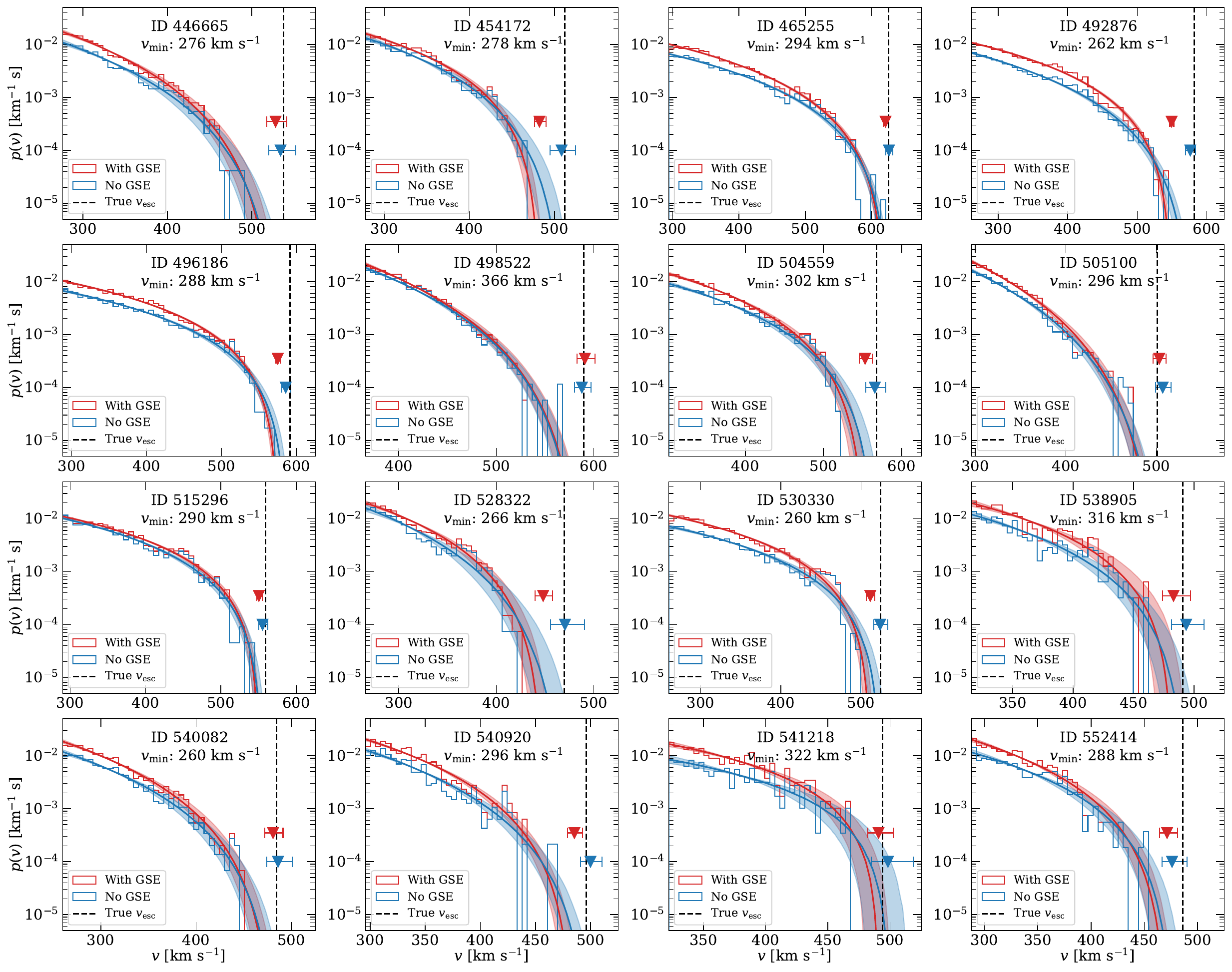}
   \caption{The comparison between the best fits and velocity distributions of high-velocity tails for the ``All sample'' of 16 final MW analogs, with the true $\vesc$ (black lines) and the fitted $\vesc$ (scatter point with error bars) marked. Red and blue represent the ``with GSE'' and the ``no GSE'' scenarios, respectively. We present only the best-fit results at our fiducial $\vmin$. }
   \label{fig:pdf_all}
   \vspace{1.5em}
\end{figure*}

\begin{figure*}[htbp]
   \centering
   \includegraphics[width=0.65\textwidth]{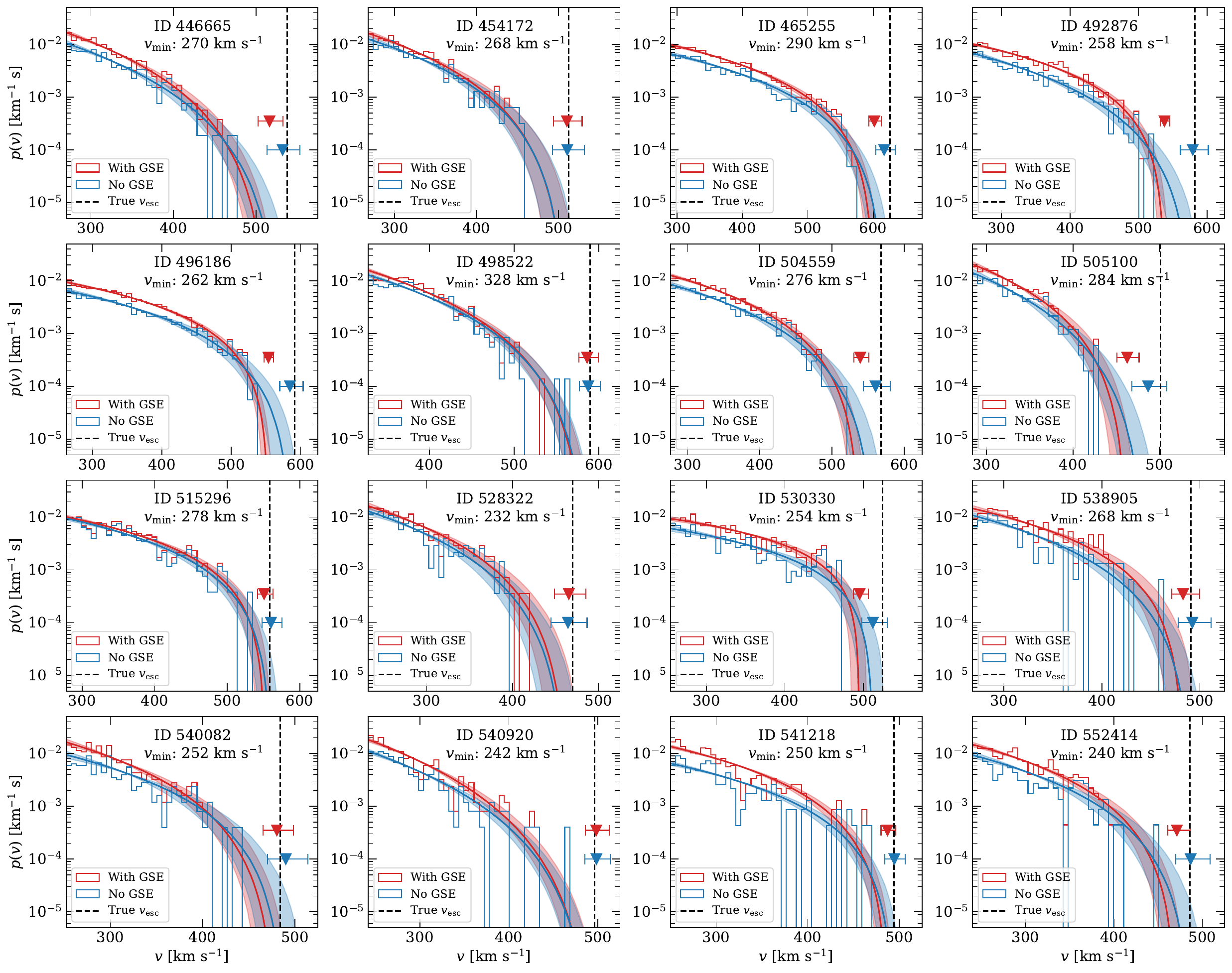}
   \caption{Similar to Figure~\ref{fig:pdf_all}, but for the ``LAMOST sample''.}
   \label{fig:pdf_lamost}
\end{figure*}

Figure~\ref{fig:pdf_all} and Figure~\ref{fig:pdf_lamost} show $\vesc$ fitting results for the 16 final MW analogs with reliable fits, corresponding to the ``All sample'' and ``LAMOST sample'', respectively.
As we can see, these galaxies exhibit relatively smooth and continuous high-velocity tails and can be well described by a single power law.

\bibliography{ref}{}

@ARTICLE{Matsuno2021,
       author = {{Matsuno}, Tadafumi and {Hirai}, Yutaka and {Tarumi}, Yuta and {Hotokezaka}, Kenta and {Tanaka}, Masaomi and {Helmi}, Amina},
        title = "{R-process enhancements of Gaia-Enceladus in GALAH DR3}",
      journal = {\aap},
         year = 2021,
        month = jun,
       volume = {650},
          eid = {A110},
        pages = {A110},
          doi = {10.1051/0004-6361/202040227},
archivePrefix = {arXiv},
       eprint = {2101.07791},
 primaryClass = {astro-ph.GA},
       adsurl = {https://ui.adsabs.harvard.edu/abs/2021A&A...650A.110M}
}

@ARTICLE{Donlon2024,
       author = {{Donlon}, Thomas and {Newberg}, Heidi Jo and {Sanderson}, Robyn and {Bregou}, Emily and {Horta}, Danny and {Arora}, Arpit and {Panithanpaisal}, Nondh},
        title = "{The debris of the 'last major merger' is dynamically young}",
      journal = {\mnras},
         year = 2024,
        month = jun,
       volume = {531},
       number = {1},
        pages = {1422-1439},
          doi = {10.1093/mnras/stae1264},
archivePrefix = {arXiv},
       eprint = {2310.09376},
 primaryClass = {astro-ph.GA},
       adsurl = {https://ui.adsabs.harvard.edu/abs/2024MNRAS.531.1422D}
}

@ARTICLE{Orkney2023,
       author = {{Orkney}, Matthew D.~A. and {Laporte}, Chervin F.~P. and {Grand}, Robert J.~J. and {G{\'o}mez}, Facundo A. and {van de Voort}, Freeke and {Fattahi}, Azadeh and {Marinacci}, Federico and {Pakmor}, R{\"u}diger and {Fragkoudi}, Francesca and {Springel}, Volker},
        title = "{Exploring the diversity and similarity of radially anisotropic Milky Way-like stellar haloes: implications for disrupted dwarf galaxy searches}",
      journal = {\mnras},
         year = 2023,
        month = oct,
       volume = {525},
       number = {1},
        pages = {683-705},
          doi = {10.1093/mnras/stad2361},
archivePrefix = {arXiv},
       eprint = {2303.02147},
 primaryClass = {astro-ph.GA},
       adsurl = {https://ui.adsabs.harvard.edu/abs/2023MNRAS.525..683O}
}

@ARTICLE{Lovell2018,
       author = {{Lovell}, Mark R. and {Pillepich}, Annalisa and {Genel}, Shy and {Nelson}, Dylan and {Springel}, Volker and {Pakmor}, R{\"u}diger and {Marinacci}, Federico and {Weinberger}, Rainer and {Torrey}, Paul and {Vogelsberger}, Mark and {Alabi}, Adebusola and {Hernquist}, Lars},
        title = "{The fraction of dark matter within galaxies from the IllustrisTNG simulations}",
      journal = {\mnras},
         year = 2018,
        month = dec,
       volume = {481},
       number = {2},
        pages = {1950-1975},
          doi = {10.1093/mnras/sty2339},
archivePrefix = {arXiv},
       eprint = {1801.10170},
 primaryClass = {astro-ph.GA},
       adsurl = {https://ui.adsabs.harvard.edu/abs/2018MNRAS.481.1950L}
}

@ARTICLE{Navarro1996,
       author = {{Navarro}, Julio F. and {Frenk}, Carlos S. and {White}, Simon D.~M.},
        title = "{The Structure of Cold Dark Matter Halos}",
      journal = {\apj},
         year = 1996,
        month = may,
       volume = {462},
        pages = {563},
          doi = {10.1086/177173},
archivePrefix = {arXiv},
       eprint = {astro-ph/9508025},
 primaryClass = {astro-ph},
       adsurl = {https://ui.adsabs.harvard.edu/abs/1996ApJ...462..563N}
}

@misc{Nelson2021,
  title = {Illustris[{{TNG}}] Public Data Release: {{Python}} Scripts},
  author = {Nelson, Dylan},
  year = {2021},
  journal = {GitHub repository},
  url = {https://github.com/illustristng/illustris_python}
}

@article{Hunter2007,
  title = {Matplotlib: {{A 2D Graphics Environment}}},
  shorttitle = {Matplotlib},
  author = {Hunter, John D.},
  year = {2007},
  month = may,
  journal = {Comput. Sci. Eng.},
  volume = {9},
  number = {3},
  pages = {90--95},
  doi = {10.1109/MCSE.2007.55},
  url = {https://ui.adsabs.harvard.edu/abs/2007CSE.....9...90H}
}

@ARTICLE{scipy,
  author  = {Virtanen, Pauli and Gommers, Ralf and Oliphant, Travis E. and
            Haberland, Matt and Reddy, Tyler and Cournapeau, David and
            Burovski, Evgeni and Peterson, Pearu and Weckesser, Warren and
            Bright, Jonathan and {van der Walt}, St{\'e}fan J. and
            Brett, Matthew and Wilson, Joshua and Millman, K. Jarrod and
            Mayorov, Nikolay and Nelson, Andrew R. J. and Jones, Eric and
            Kern, Robert and Larson, Eric and Carey, C J and
            Polat, {\.I}lhan and Feng, Yu and Moore, Eric W. and
            {VanderPlas}, Jake and Laxalde, Denis and Perktold, Josef and
            Cimrman, Robert and Henriksen, Ian and Quintero, E. A. and
            Harris, Charles R. and Archibald, Anne M. and
            Ribeiro, Ant{\^o}nio H. and Pedregosa, Fabian and
            {van Mulbregt}, Paul and {SciPy 1.0 Contributors}},
  title   = {{{SciPy} 1.0: Fundamental Algorithms for Scientific
            Computing in Python}},
  journal = {Nature Methods},
  year    = {2020},
  volume  = {17},
  pages   = {261--272},
  adsurl  = {https://rdcu.be/b08Wh},
  doi     = {10.1038/s41592-019-0686-2},
}

@Article{numpy,
 title         = {Array programming with {NumPy}},
 author        = {Charles R. Harris and K. Jarrod Millman and St{\'{e}}fan J.
                 van der Walt and Ralf Gommers and Pauli Virtanen and David
                 Cournapeau and Eric Wieser and Julian Taylor and Sebastian
                 Berg and Nathaniel J. Smith and Robert Kern and Matti Picus
                 and Stephan Hoyer and Marten H. van Kerkwijk and Matthew
                 Brett and Allan Haldane and Jaime Fern{\'{a}}ndez del
                 R{\'{i}}o and Mark Wiebe and Pearu Peterson and Pierre
                 G{\'{e}}rard-Marchant and Kevin Sheppard and Tyler Reddy and
                 Warren Weckesser and Hameer Abbasi and Christoph Gohlke and
                 Travis E. Oliphant},
 year          = {2020},
 month         = sep,
 journal       = {Nature},
 volume        = {585},
 number        = {7825},
 pages         = {357--362},
 doi           = {10.1038/s41586-020-2649-2},
 publisher     = {Springer Science and Business Media {LLC}},
 url           = {https://doi.org/10.1038/s41586-020-2649-2}
}

@article{Astropy2013,
Adsurl = {http://adsabs.harvard.edu/abs/2013A%26A...558A..33A},
Archiveprefix = {arXiv},
Author = {{Astropy Collaboration} and {Robitaille}, T.~P. and {Tollerud}, E.~J. and {Greenfield}, P. and {Droettboom}, M. and {Bray}, E. and {Aldcroft}, T. and {Davis}, M. and {Ginsburg}, A. and {Price-Whelan}, A.~M. and {Kerzendorf}, W.~E. and {Conley}, A. and {Crighton}, N. and {Barbary}, K. and {Muna}, D. and {Ferguson}, H. and {Grollier}, F. and {Parikh}, M.~M. and {Nair}, P.~H. and {Unther}, H.~M. and {Deil}, C. and {Woillez}, J. and {Conseil}, S. and {Kramer}, R. and {Turner}, J.~E.~H. and {Singer}, L. and {Fox}, R. and {Weaver}, B.~A. and {Zabalza}, V. and {Edwards}, Z.~I. and {Azalee Bostroem}, K. and {Burke}, D.~J. and {Casey}, A.~R. and {Crawford}, S.~M. and {Dencheva}, N. and {Ely}, J. and {Jenness}, T. and {Labrie}, K. and {Lim}, P.~L. and {Pierfederici}, F. and {Pontzen}, A. and {Ptak}, A. and {Refsdal}, B. and {Servillat}, M. and {Streicher}, O.},
Doi = {10.1051/0004-6361/201322068},
Eid = {A33},
Eprint = {1307.6212},
Journal = {\aap},
Month = oct,
Pages = {A33},
Primaryclass = {astro-ph.IM},
Title = {{Astropy: A community Python package for astronomy}},
Volume = 558,
Year = 2013}

@ARTICLE{Astropy2018,
       author = {{Astropy Collaboration} and {Price-Whelan}, A.~M. and
         {Sip{\H{o}}cz}, B.~M. and {G{\"u}nther}, H.~M. and {Lim}, P.~L. and
         {Crawford}, S.~M. and {Conseil}, S. and {Shupe}, D.~L. and
         {Craig}, M.~W. and {Dencheva}, N. and {Ginsburg}, A. and {Vand
        erPlas}, J.~T. and {Bradley}, L.~D. and {P{\'e}rez-Su{\'a}rez}, D. and
         {de Val-Borro}, M. and {Aldcroft}, T.~L. and {Cruz}, K.~L. and
         {Robitaille}, T.~P. and {Tollerud}, E.~J. and {Ardelean}, C. and
         {Babej}, T. and {Bach}, Y.~P. and {Bachetti}, M. and {Bakanov}, A.~V. and
         {Bamford}, S.~P. and {Barentsen}, G. and {Barmby}, P. and
         {Baumbach}, A. and {Berry}, K.~L. and {Biscani}, F. and {Boquien}, M. and
         {Bostroem}, K.~A. and {Bouma}, L.~G. and {Brammer}, G.~B. and
         {Bray}, E.~M. and {Breytenbach}, H. and {Buddelmeijer}, H. and
         {Burke}, D.~J. and {Calderone}, G. and {Cano Rodr{\'\i}guez}, J.~L. and
         {Cara}, M. and {Cardoso}, J.~V.~M. and {Cheedella}, S. and {Copin}, Y. and
         {Corrales}, L. and {Crichton}, D. and {D'Avella}, D. and {Deil}, C. and
         {Depagne}, {\'E}. and {Dietrich}, J.~P. and {Donath}, A. and
         {Droettboom}, M. and {Earl}, N. and {Erben}, T. and {Fabbro}, S. and
         {Ferreira}, L.~A. and {Finethy}, T. and {Fox}, R.~T. and
         {Garrison}, L.~H. and {Gibbons}, S.~L.~J. and {Goldstein}, D.~A. and
         {Gommers}, R. and {Greco}, J.~P. and {Greenfield}, P. and
         {Groener}, A.~M. and {Grollier}, F. and {Hagen}, A. and {Hirst}, P. and
         {Homeier}, D. and {Horton}, A.~J. and {Hosseinzadeh}, G. and {Hu}, L. and
         {Hunkeler}, J.~S. and {Ivezi{\'c}}, {\v{Z}}. and {Jain}, A. and
         {Jenness}, T. and {Kanarek}, G. and {Kendrew}, S. and {Kern}, N.~S. and
         {Kerzendorf}, W.~E. and {Khvalko}, A. and {King}, J. and {Kirkby}, D. and
         {Kulkarni}, A.~M. and {Kumar}, A. and {Lee}, A. and {Lenz}, D. and
         {Littlefair}, S.~P. and {Ma}, Z. and {Macleod}, D.~M. and
         {Mastropietro}, M. and {McCully}, C. and {Montagnac}, S. and
         {Morris}, B.~M. and {Mueller}, M. and {Mumford}, S.~J. and {Muna}, D. and
         {Murphy}, N.~A. and {Nelson}, S. and {Nguyen}, G.~H. and
         {Ninan}, J.~P. and {N{\"o}the}, M. and {Ogaz}, S. and {Oh}, S. and
         {Parejko}, J.~K. and {Parley}, N. and {Pascual}, S. and {Patil}, R. and
         {Patil}, A.~A. and {Plunkett}, A.~L. and {Prochaska}, J.~X. and
         {Rastogi}, T. and {Reddy Janga}, V. and {Sabater}, J. and
         {Sakurikar}, P. and {Seifert}, M. and {Sherbert}, L.~E. and
         {Sherwood-Taylor}, H. and {Shih}, A.~Y. and {Sick}, J. and
         {Silbiger}, M.~T. and {Singanamalla}, S. and {Singer}, L.~P. and
         {Sladen}, P.~H. and {Sooley}, K.~A. and {Sornarajah}, S. and
         {Streicher}, O. and {Teuben}, P. and {Thomas}, S.~W. and
         {Tremblay}, G.~R. and {Turner}, J.~E.~H. and {Terr{\'o}n}, V. and
         {van Kerkwijk}, M.~H. and {de la Vega}, A. and {Watkins}, L.~L. and
         {Weaver}, B.~A. and {Whitmore}, J.~B. and {Woillez}, J. and
         {Zabalza}, V. and {Astropy Contributors}},
        title = "{The Astropy Project: Building an Open-science Project and Status of the v2.0 Core Package}",
      journal = {\aj},
         year = 2018,
        month = sep,
       volume = {156},
       number = {3},
          eid = {123},
        pages = {123},
          doi = {10.3847/1538-3881/aabc4f},
archivePrefix = {arXiv},
       eprint = {1801.02634},
 primaryClass = {astro-ph.IM},
       adsurl = {https://ui.adsabs.harvard.edu/abs/2018AJ....156..123A}
}

@ARTICLE{Astropy2022,
       author = {{Astropy Collaboration} and {Price-Whelan}, Adrian M. and {Lim}, Pey Lian and {Earl}, Nicholas and {Starkman}, Nathaniel and {Bradley}, Larry and {Shupe}, David L. and {Patil}, Aarya A. and {Corrales}, Lia and {Brasseur}, C.~E. and {N{\"o}the}, Maximilian and {Donath}, Axel and {Tollerud}, Erik and {Morris}, Brett M. and {Ginsburg}, Adam and {Vaher}, Eero and {Weaver}, Benjamin A. and {Tocknell}, James and {Jamieson}, William and {van Kerkwijk}, Marten H. and {Robitaille}, Thomas P. and {Merry}, Bruce and {Bachetti}, Matteo and {G{\"u}nther}, H. Moritz and {Aldcroft}, Thomas L. and {Alvarado-Montes}, Jaime A. and {Archibald}, Anne M. and {B{\'o}di}, Attila and {Bapat}, Shreyas and {Barentsen}, Geert and {Baz{\'a}n}, Juanjo and {Biswas}, Manish and {Boquien}, M{\'e}d{\'e}ric and {Burke}, D.~J. and {Cara}, Daria and {Cara}, Mihai and {Conroy}, Kyle E. and {Conseil}, Simon and {Craig}, Matthew W. and {Cross}, Robert M. and {Cruz}, Kelle L. and {D'Eugenio}, Francesco and {Dencheva}, Nadia and {Devillepoix}, Hadrien A.~R. and {Dietrich}, J{\"o}rg P. and {Eigenbrot}, Arthur Davis and {Erben}, Thomas and {Ferreira}, Leonardo and {Foreman-Mackey}, Daniel and {Fox}, Ryan and {Freij}, Nabil and {Garg}, Suyog and {Geda}, Robel and {Glattly}, Lauren and {Gondhalekar}, Yash and {Gordon}, Karl D. and {Grant}, David and {Greenfield}, Perry and {Groener}, Austen M. and {Guest}, Steve and {Gurovich}, Sebastian and {Handberg}, Rasmus and {Hart}, Akeem and {Hatfield-Dodds}, Zac and {Homeier}, Derek and {Hosseinzadeh}, Griffin and {Jenness}, Tim and {Jones}, Craig K. and {Joseph}, Prajwel and {Kalmbach}, J. Bryce and {Karamehmetoglu}, Emir and {Ka{\l}uszy{\'n}ski}, Miko{\l}aj and {Kelley}, Michael S.~P. and {Kern}, Nicholas and {Kerzendorf}, Wolfgang E. and {Koch}, Eric W. and {Kulumani}, Shankar and {Lee}, Antony and {Ly}, Chun and {Ma}, Zhiyuan and {MacBride}, Conor and {Maljaars}, Jakob M. and {Muna}, Demitri and {Murphy}, N.~A. and {Norman}, Henrik and {O'Steen}, Richard and {Oman}, Kyle A. and {Pacifici}, Camilla and {Pascual}, Sergio and {Pascual-Granado}, J. and {Patil}, Rohit R. and {Perren}, Gabriel I. and {Pickering}, Timothy E. and {Rastogi}, Tanuj and {Roulston}, Benjamin R. and {Ryan}, Daniel F. and {Rykoff}, Eli S. and {Sabater}, Jose and {Sakurikar}, Parikshit and {Salgado}, Jes{\'u}s and {Sanghi}, Aniket and {Saunders}, Nicholas and {Savchenko}, Volodymyr and {Schwardt}, Ludwig and {Seifert-Eckert}, Michael and {Shih}, Albert Y. and {Jain}, Anany Shrey and {Shukla}, Gyanendra and {Sick}, Jonathan and {Simpson}, Chris and {Singanamalla}, Sudheesh and {Singer}, Leo P. and {Singhal}, Jaladh and {Sinha}, Manodeep and {Sip{\H{o}}cz}, Brigitta M. and {Spitler}, Lee R. and {Stansby}, David and {Streicher}, Ole and {{\v{S}}umak}, Jani and {Swinbank}, John D. and {Taranu}, Dan S. and {Tewary}, Nikita and {Tremblay}, Grant R. and {de Val-Borro}, Miguel and {Van Kooten}, Samuel J. and {Vasovi{\'c}}, Zlatan and {Verma}, Shresth and {de Miranda Cardoso}, Jos{\'e} Vin{\'\i}cius and {Williams}, Peter K.~G. and {Wilson}, Tom J. and {Winkel}, Benjamin and {Wood-Vasey}, W.~M. and {Xue}, Rui and {Yoachim}, Peter and {Zhang}, Chen and {Zonca}, Andrea and {Astropy Project Contributors}},
        title = "{The Astropy Project: Sustaining and Growing a Community-oriented Open-source Project and the Latest Major Release (v5.0) of the Core Package}",
      journal = {\apj},
         year = 2022,
        month = aug,
       volume = {935},
       number = {2},
          eid = {167},
        pages = {167},
          doi = {10.3847/1538-4357/ac7c74},
archivePrefix = {arXiv},
       eprint = {2206.14220},
 primaryClass = {astro-ph.IM},
       adsurl = {https://ui.adsabs.harvard.edu/abs/2022ApJ...935..167A}
}

@ARTICLE{Juric2008,
       author = {{Juri{\'c}}, Mario and {Ivezi{\'c}}, {\v{Z}}eljko and {Brooks}, Alyson and {Lupton}, Robert H. and {Schlegel}, David and {Finkbeiner}, Douglas and {Padmanabhan}, Nikhil and {Bond}, Nicholas and {Sesar}, Branimir and {Rockosi}, Constance M. and {Knapp}, Gillian R. and {Gunn}, James E. and {Sumi}, Takahiro and {Schneider}, Donald P. and {Barentine}, J.~C. and {Brewington}, Howard J. and {Brinkmann}, J. and {Fukugita}, Masataka and {Harvanek}, Michael and {Kleinman}, S.~J. and {Krzesinski}, Jurek and {Long}, Dan and {Neilsen}, Jr., Eric H. and {Nitta}, Atsuko and {Snedden}, Stephanie A. and {York}, Donald G.},
        title = "{The Milky Way Tomography with SDSS. I. Stellar Number Density Distribution}",
      journal = {\apj},
         year = 2008,
        month = feb,
       volume = {673},
       number = {2},
        pages = {864-914},
          doi = {10.1086/523619},
archivePrefix = {arXiv},
       eprint = {astro-ph/0510520},
 primaryClass = {astro-ph},
       adsurl = {https://ui.adsabs.harvard.edu/abs/2008ApJ...673..864J}
}

@ARTICLE{Hammer2007,
       author = {{Hammer}, F. and {Puech}, M. and {Chemin}, L. and {Flores}, H. and {Lehnert}, M.~D.},
        title = "{The Milky Way, an Exceptionally Quiet Galaxy: Implications for the Formation of Spiral Galaxies}",
      journal = {\apj},
         year = 2007,
        month = jun,
       volume = {662},
       number = {1},
        pages = {322-334},
          doi = {10.1086/516727},
archivePrefix = {arXiv},
       eprint = {astro-ph/0702585},
 primaryClass = {astro-ph},
       adsurl = {https://ui.adsabs.harvard.edu/abs/2007ApJ...662..322H}
}

@ARTICLE{Bovy2013,
       author = {{Bovy}, Jo and {Rix}, Hans-Walter},
        title = "{A Direct Dynamical Measurement of the Milky Way's Disk Surface Density Profile, Disk Scale Length, and Dark Matter Profile at 4 kpc <\raisebox{-0.5ex}\textasciitilde R <\raisebox{-0.5ex}\textasciitilde 9 kpc}",
      journal = {\apj},
         year = 2013,
        month = dec,
       volume = {779},
       number = {2},
          eid = {115},
        pages = {115},
          doi = {10.1088/0004-637X/779/2/115},
archivePrefix = {arXiv},
       eprint = {1309.0809},
 primaryClass = {astro-ph.GA},
       adsurl = {https://ui.adsabs.harvard.edu/abs/2013ApJ...779..115B}
}

@ARTICLE{Lane2022,
       author = {{Lane}, James M.~M. and {Bovy}, Jo and {Mackereth}, J. Ted},
        title = "{The kinematic properties of Milky Way stellar halo populations}",
      journal = {\mnras},
         year = 2022,
        month = mar,
       volume = {510},
       number = {4},
        pages = {5119-5141},
          doi = {10.1093/mnras/stab3755},
archivePrefix = {arXiv},
       eprint = {2106.09699},
 primaryClass = {astro-ph.GA},
       adsurl = {https://ui.adsabs.harvard.edu/abs/2022MNRAS.510.5119L}
}

@ARTICLE{Lane2023,
       author = {{Lane}, James M.~M. and {Bovy}, Jo and {Mackereth}, J. Ted},
        title = "{The stellar mass of the Gaia-Sausage/Enceladus accretion remnant}",
      journal = {\mnras},
         year = 2023,
        month = nov,
       volume = {526},
       number = {1},
        pages = {1209-1234},
          doi = {10.1093/mnras/stad2834},
archivePrefix = {arXiv},
       eprint = {2306.03084},
 primaryClass = {astro-ph.GA},
       adsurl = {https://ui.adsabs.harvard.edu/abs/2023MNRAS.526.1209L}
}

@ARTICLE{Carrillo2024,
       author = {{Carrillo}, Andreia and {Deason}, Alis J. and {Fattahi}, Azadeh and {Callingham}, Thomas M. and {Grand}, Robert J.~J.},
        title = "{Can we really pick and choose? Benchmarking various selections of Gaia Enceladus/Sausage stars in observations with simulations}",
      journal = {\mnras},
         year = 2024,
        month = jan,
       volume = {527},
       number = {2},
        pages = {2165-2184},
          doi = {10.1093/mnras/stad3274},
archivePrefix = {arXiv},
       eprint = {2306.00770},
 primaryClass = {astro-ph.GA},
       adsurl = {https://ui.adsabs.harvard.edu/abs/2024MNRAS.527.2165C}
}

@ARTICLE{Prudil2022,
       author = {{Prudil}, Z. and {Koch-Hansen}, A.~J. and {Lemasle}, B. and {Grebel}, E.~K. and {Marchetti}, T. and {Hansen}, C.~J. and {Crestani}, J. and {Braga}, V.~F. and {Bono}, G. and {Chaboyer}, B. and {Fabrizio}, M. and {Dall'Ora}, M. and {Mart{\'\i}nez-V{\'a}zquez}, C.~E.},
        title = "{Milky Way archaeology using RR Lyrae and type II Cepheids. II. High-velocity RR Lyrae stars and Milky Way mass}",
      journal = {\aap},
         year = 2022,
        month = aug,
       volume = {664},
          eid = {A148},
        pages = {A148},
          doi = {10.1051/0004-6361/202142251},
archivePrefix = {arXiv},
       eprint = {2206.00417},
 primaryClass = {astro-ph.GA},
       adsurl = {https://ui.adsabs.harvard.edu/abs/2022A&A...664A.148P}
}

@ARTICLE{Necib2019,
       author = {{Necib}, Lina and {Lisanti}, Mariangela and {Belokurov}, Vasily},
        title = "{Inferred Evidence for Dark Matter Kinematic Substructure with SDSS-Gaia}",
      journal = {\apj},
         year = 2019,
        month = mar,
       volume = {874},
       number = {1},
          eid = {3},
        pages = {3},
          doi = {10.3847/1538-4357/ab095b},
archivePrefix = {arXiv},
       eprint = {1807.02519},
 primaryClass = {astro-ph.GA},
       adsurl = {https://ui.adsabs.harvard.edu/abs/2019ApJ...874....3N}
}

@ARTICLE{Feuillet2020,
       author = {{Feuillet}, Diane K. and {Feltzing}, Sofia and {Sahlholdt}, Christian L. and {Casagrande}, Luca},
        title = "{The SkyMapper-Gaia RVS view of the Gaia-Enceladus-Sausage - an investigation of the metallicity and mass of the Milky Way's last major merger}",
      journal = {\mnras},
         year = 2020,
        month = sep,
       volume = {497},
       number = {1},
        pages = {109-124},
          doi = {10.1093/mnras/staa1888},
archivePrefix = {arXiv},
       eprint = {2003.11039},
 primaryClass = {astro-ph.GA},
       adsurl = {https://ui.adsabs.harvard.edu/abs/2020MNRAS.497..109F}
}

@ARTICLE{Hunt2025,
       author = {{Hunt}, Jason A.~S. and {Vasiliev}, Eugene},
        title = "{Milky Way dynamics in light of Gaia}",
      journal = {\nar},
         year = 2025,
        month = jun,
       volume = {100},
          eid = {101721},
        pages = {101721},
          doi = {10.1016/j.newar.2024.101721},
archivePrefix = {arXiv},
       eprint = {2501.04075},
 primaryClass = {astro-ph.GA},
       adsurl = {https://ui.adsabs.harvard.edu/abs/2025NewAR.10001721H}
}

@ARTICLE{Xie2025,
       author = {{Xie}, Renjing and {Li}, Haining and {Zhang}, Ruizhi and {Wu}, Yin and {Xue}, Xiang-Xiang and {Zhao}, Gang and {Zhang}, Shi-Lin and {Xie}, Xiao-Jin},
        title = "{Main-sequence Turnoff Stars as Probes of the Ancient Galactic Relic: Chemo-dynamical Analysis of a Pilot Sample}",
      journal = {\apj},
         year = 2025,
        month = jun,
       volume = {985},
       number = {2},
          eid = {250},
        pages = {250},
          doi = {10.3847/1538-4357/adcba6},
archivePrefix = {arXiv},
       eprint = {2506.07758},
 primaryClass = {astro-ph.GA},
       adsurl = {https://ui.adsabs.harvard.edu/abs/2025ApJ...985..250X}
}

@ARTICLE{Zhang2024,
       author = {{Zhang}, Ruizhi and {Matsuno}, Tadafumi and {Li}, Haining and {Aoki}, Wako and {Xue}, Xiang-Xiang and {Suda}, Takuma and {Zhao}, Gang and {Chen}, Yuqin and {Ishigaki}, Miho N. and {Shi}, Jianrong and {Xing}, Qianfan and {Zhao}, Jingkun},
        title = "{Four-hundred Very Metal-poor Stars Studied with LAMOST and Subaru. III. Dynamically Tagged Groups and Chemodynamical Properties}",
      journal = {\apj},
         year = 2024,
        month = may,
       volume = {966},
       number = {2},
          eid = {174},
        pages = {174},
          doi = {10.3847/1538-4357/ad31a6},
archivePrefix = {arXiv},
       eprint = {2405.02978},
 primaryClass = {astro-ph.GA},
       adsurl = {https://ui.adsabs.harvard.edu/abs/2024ApJ...966..174Z}
}

@ARTICLE{Horta2023,
       author = {{Horta}, Danny and {Schiavon}, Ricardo P. and {Mackereth}, J. Ted and {Weinberg}, David H. and {Hasselquist}, Sten and {Feuillet}, Diane and {O'Connell}, Robert W. and {Anguiano}, Borja and {Allende-Prieto}, Carlos and {Beaton}, Rachael L. and {Bizyaev}, Dmitry and {Cunha}, Katia and {Geisler}, Doug and {Garc{\'\i}a-Hern{\'a}ndez}, D.~A. and {Holtzman}, Jon and {J{\"o}nsson}, Henrik and {Lane}, Richard R. and {Majewski}, Steve R. and {M{\'e}sz{\'a}ros}, Szabolcs and {Minniti}, Dante and {Nitschelm}, Christian and {Shetrone}, Matthew and {Smith}, Verne V. and {Zasowski}, Gail},
        title = "{The chemical characterization of halo substructure in the Milky Way based on APOGEE}",
      journal = {\mnras},
         year = 2023,
        month = apr,
       volume = {520},
       number = {4},
        pages = {5671-5711},
          doi = {10.1093/mnras/stac3179},
archivePrefix = {arXiv},
       eprint = {2204.04233},
 primaryClass = {astro-ph.GA},
       adsurl = {https://ui.adsabs.harvard.edu/abs/2023MNRAS.520.5671H}
}

@ARTICLE{Dodd2023,
       author = {{Dodd}, Emma and {Callingham}, Thomas M. and {Helmi}, Amina and {Matsuno}, Tadafumi and {Ruiz-Lara}, Tom{\'a}s and {Balbinot}, Eduardo and {L{\"o}vdal}, Sofie},
        title = "{Gaia DR3 view of dynamical substructure in the stellar halo near the Sun}",
      journal = {\aap},
         year = 2023,
        month = feb,
       volume = {670},
          eid = {L2},
        pages = {L2},
          doi = {10.1051/0004-6361/202244546},
archivePrefix = {arXiv},
       eprint = {2206.11248},
 primaryClass = {astro-ph.GA},
       adsurl = {https://ui.adsabs.harvard.edu/abs/2023A&A...670L...2D}
}

@ARTICLE{Lindegren2016,
       author = {{Lindegren}, L. and {Lammers}, U. and {Bastian}, U. and {Hern{\'a}ndez}, J. and {Klioner}, S. and {Hobbs}, D. and {Bombrun}, A. and {Michalik}, D. and {Ramos-Lerate}, M. and {Butkevich}, A. and {Comoretto}, G. and {Joliet}, E. and {Holl}, B. and {Hutton}, A. and {Parsons}, P. and {Steidelm{\"u}ller}, H. and {Abbas}, U. and {Altmann}, M. and {Andrei}, A. and {Anton}, S. and {Bach}, N. and {Barache}, C. and {Becciani}, U. and {Berthier}, J. and {Bianchi}, L. and {Biermann}, M. and {Bouquillon}, S. and {Bourda}, G. and {Br{\"u}semeister}, T. and {Bucciarelli}, B. and {Busonero}, D. and {Carlucci}, T. and {Casta{\~n}eda}, J. and {Charlot}, P. and {Clotet}, M. and {Crosta}, M. and {Davidson}, M. and {de Felice}, F. and {Drimmel}, R. and {Fabricius}, C. and {Fienga}, A. and {Figueras}, F. and {Fraile}, E. and {Gai}, M. and {Garralda}, N. and {Geyer}, R. and {Gonz{\'a}lez-Vidal}, J.~J. and {Guerra}, R. and {Hambly}, N.~C. and {Hauser}, M. and {Jordan}, S. and {Lattanzi}, M.~G. and {Lenhardt}, H. and {Liao}, S. and {L{\"o}ffler}, W. and {McMillan}, P.~J. and {Mignard}, F. and {Mora}, A. and {Morbidelli}, R. and {Portell}, J. and {Riva}, A. and {Sarasso}, M. and {Serraller}, I. and {Siddiqui}, H. and {Smart}, R. and {Spagna}, A. and {Stampa}, U. and {Steele}, I. and {Taris}, F. and {Torra}, J. and {van Reeven}, W. and {Vecchiato}, A. and {Zschocke}, S. and {de Bruijne}, J. and {Gracia}, G. and {Raison}, F. and {Lister}, T. and {Marchant}, J. and {Messineo}, R. and {Soffel}, M. and {Osorio}, J. and {de Torres}, A. and {O'Mullane}, W.},
        title = "{Gaia Data Release 1. Astrometry: one billion positions, two million proper motions and parallaxes}",
      journal = {\aap},
         year = 2016,
        month = nov,
       volume = {595},
          eid = {A4},
        pages = {A4},
          doi = {10.1051/0004-6361/201628714},
archivePrefix = {arXiv},
       eprint = {1609.04303},
 primaryClass = {astro-ph.GA},
       adsurl = {https://ui.adsabs.harvard.edu/abs/2016A&A...595A...4L}
}

@ARTICLE{Naidu2021,
       author = {{Naidu}, Rohan P. and {Conroy}, Charlie and {Bonaca}, Ana and {Zaritsky}, Dennis and {Weinberger}, Rainer and {Ting}, Yuan-Sen and {Caldwell}, Nelson and {Tacchella}, Sandro and {Han}, Jiwon Jesse and {Speagle}, Joshua S. and {Cargile}, Phillip A.},
        title = "{Reconstructing the Last Major Merger of the Milky Way with the H3 Survey}",
      journal = {\apj},
         year = 2021,
        month = dec,
       volume = {923},
       number = {1},
          eid = {92},
        pages = {92},
          doi = {10.3847/1538-4357/ac2d2d},
archivePrefix = {arXiv},
       eprint = {2103.03251},
 primaryClass = {astro-ph.GA},
       adsurl = {https://ui.adsabs.harvard.edu/abs/2021ApJ...923...92N}
}

@ARTICLE{Malhan2024,
       author = {{Malhan}, Khyati and {Rix}, Hans-Walter},
        title = "{Shiva and Shakti: Presumed Proto-Galactic Fragments in the Inner Milky Way}",
      journal = {\apj},
         year = 2024,
        month = apr,
       volume = {964},
       number = {2},
          eid = {104},
        pages = {104},
          doi = {10.3847/1538-4357/ad1885},
archivePrefix = {arXiv},
       eprint = {2403.13051},
 primaryClass = {astro-ph.GA},
       adsurl = {https://ui.adsabs.harvard.edu/abs/2024ApJ...964..104M}
}

@ARTICLE{Myeong2019,
       author = {{Myeong}, G.~C. and {Vasiliev}, E. and {Iorio}, G. and {Evans}, N.~W. and {Belokurov}, V.},
        title = "{Evidence for two early accretion events that built the Milky Way stellar halo}",
      journal = {\mnras},
         year = 2019,
        month = sep,
       volume = {488},
       number = {1},
        pages = {1235-1247},
          doi = {10.1093/mnras/stz1770},
archivePrefix = {arXiv},
       eprint = {1904.03185},
 primaryClass = {astro-ph.GA},
       adsurl = {https://ui.adsabs.harvard.edu/abs/2019MNRAS.488.1235M}
}

@ARTICLE{Koppelman2019,
       author = {{Koppelman}, Helmer H. and {Helmi}, Amina and {Massari}, Davide and {Roelenga}, Sebastian and {Bastian}, Ulrich},
        title = "{Characterization and history of the Helmi streams with Gaia DR2}",
      journal = {\aap},
         year = 2019,
        month = may,
       volume = {625},
          eid = {A5},
        pages = {A5},
          doi = {10.1051/0004-6361/201834769},
archivePrefix = {arXiv},
       eprint = {1812.00846},
 primaryClass = {astro-ph.GA},
       adsurl = {https://ui.adsabs.harvard.edu/abs/2019A&A...625A...5K}
}

@ARTICLE{Malhan2022,
       author = {{Malhan}, Khyati},
        title = "{A New Member of the Milky Way's Family Tree: Characterizing the Pontus Merger of Our Galaxy}",
      journal = {\apjl},
         year = 2022,
        month = may,
       volume = {930},
       number = {1},
          eid = {L9},
        pages = {L9},
          doi = {10.3847/2041-8213/ac67da},
archivePrefix = {arXiv},
       eprint = {2204.09058},
 primaryClass = {astro-ph.GA},
       adsurl = {https://ui.adsabs.harvard.edu/abs/2022ApJ...930L...9M}
}

@ARTICLE{Yuan2020,
       author = {{Yuan}, Zhen and {Myeong}, G.~C. and {Beers}, Timothy C. and {Evans}, N.~W. and {Lee}, Young Sun and {Banerjee}, Projjwal and {Gudin}, Dmitrii and {Hattori}, Kohei and {Li}, Haining and {Matsuno}, Tadafumi and {Placco}, Vinicius M. and {Smith}, M.~C. and {Whitten}, Devin D. and {Zhao}, Gang},
        title = "{Dynamical Relics of the Ancient Galactic Halo}",
      journal = {\apj},
         year = 2020,
        month = mar,
       volume = {891},
       number = {1},
          eid = {39},
        pages = {39},
          doi = {10.3847/1538-4357/ab6ef7},
archivePrefix = {arXiv},
       eprint = {1910.07538},
 primaryClass = {astro-ph.GA},
       adsurl = {https://ui.adsabs.harvard.edu/abs/2020ApJ...891...39Y}
}

@ARTICLE{Matsuno2019,
       author = {{Matsuno}, Tadafumi and {Aoki}, Wako and {Suda}, Takuma},
        title = "{Origin of the Excess of High-energy Retrograde Stars in the Galactic Halo}",
      journal = {\apjl},
         year = 2019,
        month = apr,
       volume = {874},
       number = {2},
          eid = {L35},
        pages = {L35},
          doi = {10.3847/2041-8213/ab0ec0},
archivePrefix = {arXiv},
       eprint = {1903.09456},
 primaryClass = {astro-ph.GA},
       adsurl = {https://ui.adsabs.harvard.edu/abs/2019ApJ...874L..35M}
}

@ARTICLE{Naidu2020,
       author = {{Naidu}, Rohan P. and {Conroy}, Charlie and {Bonaca}, Ana and {Johnson}, Benjamin D. and {Ting}, Yuan-Sen and {Caldwell}, Nelson and {Zaritsky}, Dennis and {Cargile}, Phillip A.},
        title = "{Evidence from the H3 Survey That the Stellar Halo Is Entirely Comprised of Substructure}",
      journal = {\apj},
         year = 2020,
        month = sep,
       volume = {901},
       number = {1},
          eid = {48},
        pages = {48},
          doi = {10.3847/1538-4357/abaef4},
archivePrefix = {arXiv},
       eprint = {2006.08625},
 primaryClass = {astro-ph.GA},
       adsurl = {https://ui.adsabs.harvard.edu/abs/2020ApJ...901...48N}
}

@ARTICLE{Haywood2018,
       author = {{Haywood}, M. and {Di Matteo}, P. and {Lehnert}, M.~D. and {Snaith}, O. and {Khoperskov}, S. and {G{\'o}mez}, A.},
        title = "{In Disguise or Out of Reach: First Clues about In Situ and Accreted Stars in the Stellar Halo of the Milky Way from Gaia DR2}",
      journal = {\apj},
         year = 2018,
        month = aug,
       volume = {863},
       number = {2},
          eid = {113},
        pages = {113},
          doi = {10.3847/1538-4357/aad235},
archivePrefix = {arXiv},
       eprint = {1805.02617},
 primaryClass = {astro-ph.GA},
       adsurl = {https://ui.adsabs.harvard.edu/abs/2018ApJ...863..113H}
}

@ARTICLE{Helmi2018,
       author = {{Helmi}, Amina and {Babusiaux}, Carine and {Koppelman}, Helmer H. and {Massari}, Davide and {Veljanoski}, Jovan and {Brown}, Anthony G.~A.},
        title = "{The merger that led to the formation of the Milky Way's inner stellar halo and thick disk}",
      journal = {\nat},
         year = 2018,
        month = oct,
       volume = {563},
       number = {7729},
        pages = {85-88},
          doi = {10.1038/s41586-018-0625-x},
archivePrefix = {arXiv},
       eprint = {1806.06038},
 primaryClass = {astro-ph.GA},
       adsurl = {https://ui.adsabs.harvard.edu/abs/2018Natur.563...85H}
}

@ARTICLE{Belokurov2018,
       author = {{Belokurov}, V. and {Erkal}, D. and {Evans}, N.~W. and {Koposov}, S.~E. and {Deason}, A.~J.},
        title = "{Co-formation of the disc and the stellar halo}",
      journal = {\mnras},
         year = 2018,
        month = jul,
       volume = {478},
       number = {1},
        pages = {611-619},
          doi = {10.1093/mnras/sty982},
archivePrefix = {arXiv},
       eprint = {1802.03414},
 primaryClass = {astro-ph.GA},
       adsurl = {https://ui.adsabs.harvard.edu/abs/2018MNRAS.478..611B}
}

@ARTICLE{Gaia2023,
       author = {{Gaia Collaboration} and {Vallenari}, A. and {Brown}, A.~G.~A. and {Prusti}, T. and {de Bruijne}, J.~H.~J. and {Arenou}, F. and {Babusiaux}, C. and {Biermann}, M. and {Creevey}, O.~L. and {Ducourant}, C. and {Evans}, D.~W. and {Eyer}, L. and {Guerra}, R. and {Hutton}, A. and {Jordi}, C. and {Klioner}, S.~A. and {Lammers}, U.~L. and {Lindegren}, L. and {Luri}, X. and {Mignard}, F. and {Panem}, C. and {Pourbaix}, D. and {Randich}, S. and {Sartoretti}, P. and {Soubiran}, C. and {Tanga}, P. and {Walton}, N.~A. and {Bailer-Jones}, C.~A.~L. and {Bastian}, U. and {Drimmel}, R. and {Jansen}, F. and {Katz}, D. and {Lattanzi}, M.~G. and {van Leeuwen}, F. and {Bakker}, J. and {Cacciari}, C. and {Casta{\~n}eda}, J. and {De Angeli}, F. and {Fabricius}, C. and {Fouesneau}, M. and {Fr{\'e}mat}, Y. and {Galluccio}, L. and {Guerrier}, A. and {Heiter}, U. and {Masana}, E. and {Messineo}, R. and {Mowlavi}, N. and {Nicolas}, C. and {Nienartowicz}, K. and {Pailler}, F. and {Panuzzo}, P. and {Riclet}, F. and {Roux}, W. and {Seabroke}, G.~M. and {Sordo}, R. and {Th{\'e}venin}, F. and {Gracia-Abril}, G. and {Portell}, J. and {Teyssier}, D. and {Altmann}, M. and {Andrae}, R. and {Audard}, M. and {Bellas-Velidis}, I. and {Benson}, K. and {Berthier}, J. and {Blomme}, R. and {Burgess}, P.~W. and {Busonero}, D. and {Busso}, G. and {C{\'a}novas}, H. and {Carry}, B. and {Cellino}, A. and {Cheek}, N. and {Clementini}, G. and {Damerdji}, Y. and {Davidson}, M. and {de Teodoro}, P. and {Nu{\~n}ez Campos}, M. and {Delchambre}, L. and {Dell'Oro}, A. and {Esquej}, P. and {Fern{\'a}ndez-Hern{\'a}ndez}, J. and {Fraile}, E. and {Garabato}, D. and {Garc{\'\i}a-Lario}, P. and {Gosset}, E. and {Haigron}, R. and {Halbwachs}, J. -L. and {Hambly}, N.~C. and {Harrison}, D.~L. and {Hern{\'a}ndez}, J. and {Hestroffer}, D. and {Hodgkin}, S.~T. and {Holl}, B. and {Jan{\ss}en}, K. and {Jevardat de Fombelle}, G. and {Jordan}, S. and {Krone-Martins}, A. and {Lanzafame}, A.~C. and {L{\"o}ffler}, W. and {Marchal}, O. and {Marrese}, P.~M. and {Moitinho}, A. and {Muinonen}, K. and {Osborne}, P. and {Pancino}, E. and {Pauwels}, T. and {Recio-Blanco}, A. and {Reyl{\'e}}, C. and {Riello}, M. and {Rimoldini}, L. and {Roegiers}, T. and {Rybizki}, J. and {Sarro}, L.~M. and {Siopis}, C. and {Smith}, M. and {Sozzetti}, A. and {Utrilla}, E. and {van Leeuwen}, M. and {Abbas}, U. and {{\'A}brah{\'a}m}, P. and {Abreu Aramburu}, A. and {Aerts}, C. and {Aguado}, J.~J. and {Ajaj}, M. and {Aldea-Montero}, F. and {Altavilla}, G. and {{\'A}lvarez}, M.~A. and {Alves}, J. and {Anders}, F. and {Anderson}, R.~I. and {Anglada Varela}, E. and {Antoja}, T. and {Baines}, D. and {Baker}, S.~G. and {Balaguer-N{\'u}{\~n}ez}, L. and {Balbinot}, E. and {Balog}, Z. and {Barache}, C. and {Barbato}, D. and {Barros}, M. and {Barstow}, M.~A. and {Bartolom{\'e}}, S. and {Bassilana}, J. -L. and {Bauchet}, N. and {Becciani}, U. and {Bellazzini}, M. and {Berihuete}, A. and {Bernet}, M. and {Bertone}, S. and {Bianchi}, L. and {Binnenfeld}, A. and {Blanco-Cuaresma}, S. and {Blazere}, A. and {Boch}, T. and {Bombrun}, A. and {Bossini}, D. and {Bouquillon}, S. and {Bragaglia}, A. and {Bramante}, L. and {Breedt}, E. and {Bressan}, A. and {Brouillet}, N. and {Brugaletta}, E. and {Bucciarelli}, B. and {Burlacu}, A. and {Butkevich}, A.~G. and {Buzzi}, R. and {Caffau}, E. and {Cancelliere}, R. and {Cantat-Gaudin}, T. and {Carballo}, R. and {Carlucci}, T. and {Carnerero}, M.~I. and {Carrasco}, J.~M. and {Casamiquela}, L. and {Castellani}, M. and {Castro-Ginard}, A. and {Chaoul}, L. and {Charlot}, P. and {Chemin}, L. and {Chiaramida}, V. and {Chiavassa}, A. and {Chornay}, N. and {Comoretto}, G. and {Contursi}, G. and {Cooper}, W.~J. and {Cornez}, T. and {Cowell}, S. and {Crifo}, F. and {Cropper}, M. and {Crosta}, M. and {Crowley}, C. and {Dafonte}, C. and {Dapergolas}, A. and {David}, M. and {David}, P. and {de Laverny}, P. and {De Luise}, F. and {De March}, R. and {De Ridder}, J. and {de Souza}, R. and {de Torres}, A. and {del Peloso}, E.~F. and {del Pozo}, E. and {Delbo}, M. and {Delgado}, A. and {Delisle}, J. -B. and {Demouchy}, C. and {Dharmawardena}, T.~E. and {Di Matteo}, P. and {Diakite}, S. and {Diener}, C. and {Distefano}, E. and {Dolding}, C. and {Edvardsson}, B. and {Enke}, H. and {Fabre}, C. and {Fabrizio}, M. and {Faigler}, S. and {Fedorets}, G. and {Fernique}, P. and {Fienga}, A. and {Figueras}, F. and {Fournier}, Y. and {Fouron}, C. and {Fragkoudi}, F. and {Gai}, M. and {Garcia-Gutierrez}, A. and {Garcia-Reinaldos}, M. and {Garc{\'\i}a-Torres}, M. and {Garofalo}, A. and {Gavel}, A. and {Gavras}, P. and {Gerlach}, E. and {Geyer}, R. and {Giacobbe}, P. and {Gilmore}, G. and {Girona}, S. and {Giuffrida}, G. and {Gomel}, R. and {Gomez}, A. and {Gonz{\'a}lez-N{\'u}{\~n}ez}, J. and {Gonz{\'a}lez-Santamar{\'\i}a}, I. and {Gonz{\'a}lez-Vidal}, J.~J. and {Granvik}, M. and {Guillout}, P. and {Guiraud}, J. and {Guti{\'e}rrez-S{\'a}nchez}, R. and {Guy}, L.~P. and {Hatzidimitriou}, D. and {Hauser}, M. and {Haywood}, M. and {Helmer}, A. and {Helmi}, A. and {Sarmiento}, M.~H. and {Hidalgo}, S.~L. and {Hilger}, T. and {H{\l}adczuk}, N. and {Hobbs}, D. and {Holland}, G. and {Huckle}, H.~E. and {Jardine}, K. and {Jasniewicz}, G. and {Jean-Antoine Piccolo}, A. and {Jim{\'e}nez-Arranz}, {\'O}. and {Jorissen}, A. and {Juaristi Campillo}, J. and {Julbe}, F. and {Karbevska}, L. and {Kervella}, P. and {Khanna}, S. and {Kontizas}, M. and {Kordopatis}, G. and {Korn}, A.~J. and {K{\'o}sp{\'a}l}, {\'A}. and {Kostrzewa-Rutkowska}, Z. and {Kruszy{\'n}ska}, K. and {Kun}, M. and {Laizeau}, P. and {Lambert}, S. and {Lanza}, A.~F. and {Lasne}, Y. and {Le Campion}, J. -F. and {Lebreton}, Y. and {Lebzelter}, T. and {Leccia}, S. and {Leclerc}, N. and {Lecoeur-Taibi}, I. and {Liao}, S. and {Licata}, E.~L. and {Lindstr{\o}m}, H.~E.~P. and {Lister}, T.~A. and {Livanou}, E. and {Lobel}, A. and {Lorca}, A. and {Loup}, C. and {Madrero Pardo}, P. and {Magdaleno Romeo}, A. and {Managau}, S. and {Mann}, R.~G. and {Manteiga}, M. and {Marchant}, J.~M. and {Marconi}, M. and {Marcos}, J. and {Marcos Santos}, M.~M.~S. and {Mar{\'\i}n Pina}, D. and {Marinoni}, S. and {Marocco}, F. and {Marshall}, D.~J. and {Martin Polo}, L. and {Mart{\'\i}n-Fleitas}, J.~M. and {Marton}, G. and {Mary}, N. and {Masip}, A. and {Massari}, D. and {Mastrobuono-Battisti}, A. and {Mazeh}, T. and {McMillan}, P.~J. and {Messina}, S. and {Michalik}, D. and {Millar}, N.~R. and {Mints}, A. and {Molina}, D. and {Molinaro}, R. and {Moln{\'a}r}, L. and {Monari}, G. and {Mongui{\'o}}, M. and {Montegriffo}, P. and {Montero}, A. and {Mor}, R. and {Mora}, A. and {Morbidelli}, R. and {Morel}, T. and {Morris}, D. and {Muraveva}, T. and {Murphy}, C.~P. and {Musella}, I. and {Nagy}, Z. and {Noval}, L. and {Oca{\~n}a}, F. and {Ogden}, A. and {Ordenovic}, C. and {Osinde}, J.~O. and {Pagani}, C. and {Pagano}, I. and {Palaversa}, L. and {Palicio}, P.~A. and {Pallas-Quintela}, L. and {Panahi}, A. and {Payne-Wardenaar}, S. and {Pe{\~n}alosa Esteller}, X. and {Penttil{\"a}}, A. and {Pichon}, B. and {Piersimoni}, A.~M. and {Pineau}, F. -X. and {Plachy}, E. and {Plum}, G. and {Poggio}, E. and {Pr{\v{s}}a}, A. and {Pulone}, L. and {Racero}, E. and {Ragaini}, S. and {Rainer}, M. and {Raiteri}, C.~M. and {Rambaux}, N. and {Ramos}, P. and {Ramos-Lerate}, M. and {Re Fiorentin}, P. and {Regibo}, S. and {Richards}, P.~J. and {Rios Diaz}, C. and {Ripepi}, V. and {Riva}, A. and {Rix}, H. -W. and {Rixon}, G. and {Robichon}, N. and {Robin}, A.~C. and {Robin}, C. and {Roelens}, M. and {Rogues}, H.~R.~O. and {Rohrbasser}, L. and {Romero-G{\'o}mez}, M. and {Rowell}, N. and {Royer}, F. and {Ruz Mieres}, D. and {Rybicki}, K.~A. and {Sadowski}, G. and {S{\'a}ez N{\'u}{\~n}ez}, A. and {Sagrist{\`a} Sell{\'e}s}, A. and {Sahlmann}, J. and {Salguero}, E. and {Samaras}, N. and {Sanchez Gimenez}, V. and {Sanna}, N. and {Santove{\~n}a}, R. and {Sarasso}, M. and {Schultheis}, M. and {Sciacca}, E. and {Segol}, M. and {Segovia}, J.~C. and {S{\'e}gransan}, D. and {Semeux}, D. and {Shahaf}, S. and {Siddiqui}, H.~I. and {Siebert}, A. and {Siltala}, L. and {Silvelo}, A. and {Slezak}, E. and {Slezak}, I. and {Smart}, R.~L. and {Snaith}, O.~N. and {Solano}, E. and {Solitro}, F. and {Souami}, D. and {Souchay}, J. and {Spagna}, A. and {Spina}, L. and {Spoto}, F. and {Steele}, I.~A. and {Steidelm{\"u}ller}, H. and {Stephenson}, C.~A. and {S{\"u}veges}, M. and {Surdej}, J. and {Szabados}, L. and {Szegedi-Elek}, E. and {Taris}, F. and {Taylor}, M.~B. and {Teixeira}, R. and {Tolomei}, L. and {Tonello}, N. and {Torra}, F. and {Torra}, J. and {Torralba Elipe}, G. and {Trabucchi}, M. and {Tsounis}, A.~T. and {Turon}, C. and {Ulla}, A. and {Unger}, N. and {Vaillant}, M.~V. and {van Dillen}, E. and {van Reeven}, W. and {Vanel}, O. and {Vecchiato}, A. and {Viala}, Y. and {Vicente}, D. and {Voutsinas}, S. and {Weiler}, M. and {Wevers}, T. and {Wyrzykowski}, {\L}. and {Yoldas}, A. and {Yvard}, P. and {Zhao}, H. and {Zorec}, J. and {Zucker}, S. and {Zwitter}, T.},
        title = "{Gaia Data Release 3. Summary of the content and survey properties}",
      journal = {\aap},
         year = 2023,
        month = jun,
       volume = {674},
          eid = {A1},
        pages = {A1},
          doi = {10.1051/0004-6361/202243940},
archivePrefix = {arXiv},
       eprint = {2208.00211},
 primaryClass = {astro-ph.GA},
       adsurl = {https://ui.adsabs.harvard.edu/abs/2023A&A...674A...1G}
}

@ARTICLE{Steinmetz2006,
       author = {{Steinmetz}, M. and {Zwitter}, T. and {Siebert}, A. and {Watson}, F.~G. and {Freeman}, K.~C. and {Munari}, U. and {Campbell}, R. and {Williams}, M. and {Seabroke}, G.~M. and {Wyse}, R.~F.~G. and {Parker}, Q.~A. and {Bienaym{\'e}}, O. and {Roeser}, S. and {Gibson}, B.~K. and {Gilmore}, G. and {Grebel}, E.~K. and {Helmi}, A. and {Navarro}, J.~F. and {Burton}, D. and {Cass}, C.~J.~P. and {Dawe}, J.~A. and {Fiegert}, K. and {Hartley}, M. and {Russell}, K.~S. and {Saunders}, W. and {Enke}, H. and {Bailin}, J. and {Binney}, J. and {Bland-Hawthorn}, J. and {Boeche}, C. and {Dehnen}, W. and {Eisenstein}, D.~J. and {Evans}, N.~W. and {Fiorucci}, M. and {Fulbright}, J.~P. and {Gerhard}, O. and {Jauregi}, U. and {Kelz}, A. and {Mijovi{\'c}}, L. and {Minchev}, I. and {Parmentier}, G. and {Pe{\~n}arrubia}, J. and {Quillen}, A.~C. and {Read}, M.~A. and {Ruchti}, G. and {Scholz}, R. -D. and {Siviero}, A. and {Smith}, M.~C. and {Sordo}, R. and {Veltz}, L. and {Vidrih}, S. and {von Berlepsch}, R. and {Boyle}, B.~J. and {Schilbach}, E.},
        title = "{The Radial Velocity Experiment (RAVE): First Data Release}",
      journal = {\aj},
         year = 2006,
        month = oct,
       volume = {132},
       number = {4},
        pages = {1645-1668},
          doi = {10.1086/506564},
archivePrefix = {arXiv},
       eprint = {astro-ph/0606211},
 primaryClass = {astro-ph},
       adsurl = {https://ui.adsabs.harvard.edu/abs/2006AJ....132.1645S}
}

@ARTICLE{York2000,
       author = {{York}, Donald G. and {Adelman}, J. and {Anderson}, Jr., John E. and {Anderson}, Scott F. and {Annis}, James and {Bahcall}, Neta A. and {Bakken}, J.~A. and {Barkhouser}, Robert and {Bastian}, Steven and {Berman}, Eileen and {Boroski}, William N. and {Bracker}, Steve and {Briegel}, Charlie and {Briggs}, John W. and {Brinkmann}, J. and {Brunner}, Robert and {Burles}, Scott and {Carey}, Larry and {Carr}, Michael A. and {Castander}, Francisco J. and {Chen}, Bing and {Colestock}, Patrick L. and {Connolly}, A.~J. and {Crocker}, J.~H. and {Csabai}, Istv{\'a}n and {Czarapata}, Paul C. and {Davis}, John Eric and {Doi}, Mamoru and {Dombeck}, Tom and {Eisenstein}, Daniel and {Ellman}, Nancy and {Elms}, Brian R. and {Evans}, Michael L. and {Fan}, Xiaohui and {Federwitz}, Glenn R. and {Fiscelli}, Larry and {Friedman}, Scott and {Frieman}, Joshua A. and {Fukugita}, Masataka and {Gillespie}, Bruce and {Gunn}, James E. and {Gurbani}, Vijay K. and {de Haas}, Ernst and {Haldeman}, Merle and {Harris}, Frederick H. and {Hayes}, J. and {Heckman}, Timothy M. and {Hennessy}, G.~S. and {Hindsley}, Robert B. and {Holm}, Scott and {Holmgren}, Donald J. and {Huang}, Chi-hao and {Hull}, Charles and {Husby}, Don and {Ichikawa}, Shin-Ichi and {Ichikawa}, Takashi and {Ivezi{\'c}}, {\v{Z}}eljko and {Kent}, Stephen and {Kim}, Rita S.~J. and {Kinney}, E. and {Klaene}, Mark and {Kleinman}, A.~N. and {Kleinman}, S. and {Knapp}, G.~R. and {Korienek}, John and {Kron}, Richard G. and {Kunszt}, Peter Z. and {Lamb}, D.~Q. and {Lee}, B. and {Leger}, R. French and {Limmongkol}, Siriluk and {Lindenmeyer}, Carl and {Long}, Daniel C. and {Loomis}, Craig and {Loveday}, Jon and {Lucinio}, Rich and {Lupton}, Robert H. and {MacKinnon}, Bryan and {Mannery}, Edward J. and {Mantsch}, P.~M. and {Margon}, Bruce and {McGehee}, Peregrine and {McKay}, Timothy A. and {Meiksin}, Avery and {Merelli}, Aronne and {Monet}, David G. and {Munn}, Jeffrey A. and {Narayanan}, Vijay K. and {Nash}, Thomas and {Neilsen}, Eric and {Neswold}, Rich and {Newberg}, Heidi Jo and {Nichol}, R.~C. and {Nicinski}, Tom and {Nonino}, Mario and {Okada}, Norio and {Okamura}, Sadanori and {Ostriker}, Jeremiah P. and {Owen}, Russell and {Pauls}, A. George and {Peoples}, John and {Peterson}, R.~L. and {Petravick}, Donald and {Pier}, Jeffrey R. and {Pope}, Adrian and {Pordes}, Ruth and {Prosapio}, Angela and {Rechenmacher}, Ron and {Quinn}, Thomas R. and {Richards}, Gordon T. and {Richmond}, Michael W. and {Rivetta}, Claudio H. and {Rockosi}, Constance M. and {Ruthmansdorfer}, Kurt and {Sandford}, Dale and {Schlegel}, David J. and {Schneider}, Donald P. and {Sekiguchi}, Maki and {Sergey}, Gary and {Shimasaku}, Kazuhiro and {Siegmund}, Walter A. and {Smee}, Stephen and {Smith}, J. Allyn and {Snedden}, S. and {Stone}, R. and {Stoughton}, Chris and {Strauss}, Michael A. and {Stubbs}, Christopher and {SubbaRao}, Mark and {Szalay}, Alexander S. and {Szapudi}, Istvan and {Szokoly}, Gyula P. and {Thakar}, Anirudda R. and {Tremonti}, Christy and {Tucker}, Douglas L. and {Uomoto}, Alan and {Vanden Berk}, Dan and {Vogeley}, Michael S. and {Waddell}, Patrick and {Wang}, Shu-i. and {Watanabe}, Masaru and {Weinberg}, David H. and {Yanny}, Brian and {Yasuda}, Naoki and {SDSS Collaboration}},
        title = "{The Sloan Digital Sky Survey: Technical Summary}",
      journal = {\aj},
         year = 2000,
        month = sep,
       volume = {120},
       number = {3},
        pages = {1579-1587},
          doi = {10.1086/301513},
archivePrefix = {arXiv},
       eprint = {astro-ph/0006396},
 primaryClass = {astro-ph},
       adsurl = {https://ui.adsabs.harvard.edu/abs/2000AJ....120.1579Y}
}

@ARTICLE{Majewski2017,
       author = {{Majewski}, Steven R. and {Schiavon}, Ricardo P. and {Frinchaboy}, Peter M. and {Allende Prieto}, Carlos and {Barkhouser}, Robert and {Bizyaev}, Dmitry and {Blank}, Basil and {Brunner}, Sophia and {Burton}, Adam and {Carrera}, Ricardo and {Chojnowski}, S. Drew and {Cunha}, K{\'a}tia and {Epstein}, Courtney and {Fitzgerald}, Greg and {Garc{\'\i}a P{\'e}rez}, Ana E. and {Hearty}, Fred R. and {Henderson}, Chuck and {Holtzman}, Jon A. and {Johnson}, Jennifer A. and {Lam}, Charles R. and {Lawler}, James E. and {Maseman}, Paul and {M{\'e}sz{\'a}ros}, Szabolcs and {Nelson}, Matthew and {Nguyen}, Duy Coung and {Nidever}, David L. and {Pinsonneault}, Marc and {Shetrone}, Matthew and {Smee}, Stephen and {Smith}, Verne V. and {Stolberg}, Todd and {Skrutskie}, Michael F. and {Walker}, Eric and {Wilson}, John C. and {Zasowski}, Gail and {Anders}, Friedrich and {Basu}, Sarbani and {Beland}, Stephane and {Blanton}, Michael R. and {Bovy}, Jo and {Brownstein}, Joel R. and {Carlberg}, Joleen and {Chaplin}, William and {Chiappini}, Cristina and {Eisenstein}, Daniel J. and {Elsworth}, Yvonne and {Feuillet}, Diane and {Fleming}, Scott W. and {Galbraith-Frew}, Jessica and {Garc{\'\i}a}, Rafael A. and {Garc{\'\i}a-Hern{\'a}ndez}, D. An{\'\i}bal and {Gillespie}, Bruce A. and {Girardi}, L{\'e}o and {Gunn}, James E. and {Hasselquist}, Sten and {Hayden}, Michael R. and {Hekker}, Saskia and {Ivans}, Inese and {Kinemuchi}, Karen and {Klaene}, Mark and {Mahadevan}, Suvrath and {Mathur}, Savita and {Mosser}, Beno{\^\i}t and {Muna}, Demitri and {Munn}, Jeffrey A. and {Nichol}, Robert C. and {O'Connell}, Robert W. and {Parejko}, John K. and {Robin}, A.~C. and {Rocha-Pinto}, Helio and {Schultheis}, Matthias and {Serenelli}, Aldo M. and {Shane}, Neville and {Silva Aguirre}, Victor and {Sobeck}, Jennifer S. and {Thompson}, Benjamin and {Troup}, Nicholas W. and {Weinberg}, David H. and {Zamora}, Olga},
        title = "{The Apache Point Observatory Galactic Evolution Experiment (APOGEE)}",
      journal = {\aj},
         year = 2017,
        month = sep,
       volume = {154},
       number = {3},
          eid = {94},
        pages = {94},
          doi = {10.3847/1538-3881/aa784d},
archivePrefix = {arXiv},
       eprint = {1509.05420},
 primaryClass = {astro-ph.IM},
       adsurl = {https://ui.adsabs.harvard.edu/abs/2017AJ....154...94M}
}

@ARTICLE{Silva2015,
       author = {{De Silva}, G.~M. and {Freeman}, K.~C. and {Bland-Hawthorn}, J. and {Martell}, S. and {de Boer}, E. Wylie and {Asplund}, M. and {Keller}, S. and {Sharma}, S. and {Zucker}, D.~B. and {Zwitter}, T. and {Anguiano}, B. and {Bacigalupo}, C. and {Bayliss}, D. and {Beavis}, M.~A. and {Bergemann}, M. and {Campbell}, S. and {Cannon}, R. and {Carollo}, D. and {Casagrande}, L. and {Casey}, A.~R. and {Da Costa}, G. and {D'Orazi}, V. and {Dotter}, A. and {Duong}, L. and {Heger}, A. and {Ireland}, M.~J. and {Kafle}, P.~R. and {Kos}, J. and {Lattanzio}, J. and {Lewis}, G.~F. and {Lin}, J. and {Lind}, K. and {Munari}, U. and {Nataf}, D.~M. and {O'Toole}, S. and {Parker}, Q. and {Reid}, W. and {Schlesinger}, K.~J. and {Sheinis}, A. and {Simpson}, J.~D. and {Stello}, D. and {Ting}, Y. -S. and {Traven}, G. and {Watson}, F. and {Wittenmyer}, R. and {Yong}, D. and {{\v{Z}}erjal}, M.},
        title = "{The GALAH survey: scientific motivation}",
      journal = {\mnras},
         year = 2015,
        month = may,
       volume = {449},
       number = {3},
        pages = {2604-2617},
          doi = {10.1093/mnras/stv327},
archivePrefix = {arXiv},
       eprint = {1502.04767},
 primaryClass = {astro-ph.GA},
       adsurl = {https://ui.adsabs.harvard.edu/abs/2015MNRAS.449.2604D}
}

@ARTICLE{Helmi1999,
       author = {{Helmi}, Amina and {White}, Simon D.~M. and {de Zeeuw}, P. Tim and {Zhao}, Hongsheng},
        title = "{Debris streams in the solar neighbourhood as relicts from the formation of the Milky Way}",
      journal = {\nat},
         year = 1999,
        month = nov,
       volume = {402},
       number = {6757},
        pages = {53-55},
          doi = {10.1038/46980},
archivePrefix = {arXiv},
       eprint = {astro-ph/9911041},
 primaryClass = {astro-ph},
       adsurl = {https://ui.adsabs.harvard.edu/abs/1999Natur.402...53H}
}

@ARTICLE{Belokurov2006,
       author = {{Belokurov}, V. and {Zucker}, D.~B. and {Evans}, N.~W. and {Gilmore}, G. and {Vidrih}, S. and {Bramich}, D.~M. and {Newberg}, H.~J. and {Wyse}, R.~F.~G. and {Irwin}, M.~J. and {Fellhauer}, M. and {Hewett}, P.~C. and {Walton}, N.~A. and {Wilkinson}, M.~I. and {Cole}, N. and {Yanny}, B. and {Rockosi}, C.~M. and {Beers}, T.~C. and {Bell}, E.~F. and {Brinkmann}, J. and {Ivezi{\'c}}, {\v{Z}}. and {Lupton}, R.},
        title = "{The Field of Streams: Sagittarius and Its Siblings}",
      journal = {\apjl},
         year = 2006,
        month = may,
       volume = {642},
       number = {2},
        pages = {L137-L140},
          doi = {10.1086/504797},
archivePrefix = {arXiv},
       eprint = {astro-ph/0605025},
 primaryClass = {astro-ph},
       adsurl = {https://ui.adsabs.harvard.edu/abs/2006ApJ...642L.137B}
}

@ARTICLE{Bullock2005,
       author = {{Bullock}, James S. and {Johnston}, Kathryn V.},
        title = "{Tracing Galaxy Formation with Stellar Halos. I. Methods}",
      journal = {\apj},
         year = 2005,
        month = dec,
       volume = {635},
       number = {2},
        pages = {931-949},
          doi = {10.1086/497422},
archivePrefix = {arXiv},
       eprint = {astro-ph/0506467},
 primaryClass = {astro-ph},
       adsurl = {https://ui.adsabs.harvard.edu/abs/2005ApJ...635..931B}
}

@ARTICLE{Helmi2020,
       author = {{Helmi}, Amina},
        title = "{Streams, Substructures, and the Early History of the Milky Way}",
      journal = {\araa},
         year = 2020,
        month = aug,
       volume = {58},
        pages = {205-256},
          doi = {10.1146/annurev-astro-032620-021917},
archivePrefix = {arXiv},
       eprint = {2002.04340},
 primaryClass = {astro-ph.GA},
       adsurl = {https://ui.adsabs.harvard.edu/abs/2020ARA&A..58..205H}
}

@ARTICLE{Deason2024,
       author = {{Deason}, Alis J. and {Belokurov}, Vasily},
        title = "{Galactic Archaeology with Gaia}",
      journal = {\nar},
         year = 2024,
        month = dec,
       volume = {99},
          eid = {101706},
        pages = {101706},
          doi = {10.1016/j.newar.2024.101706},
archivePrefix = {arXiv},
       eprint = {2402.12443},
 primaryClass = {astro-ph.GA},
       adsurl = {https://ui.adsabs.harvard.edu/abs/2024NewAR..9901706D}
}

@ARTICLE{Press1974,
       author = {{Press}, William H. and {Schechter}, Paul},
        title = "{Formation of Galaxies and Clusters of Galaxies by Self-Similar Gravitational Condensation}",
      journal = {\apj},
         year = 1974,
        month = feb,
       volume = {187},
        pages = {425-438},
          doi = {10.1086/152650},
       adsurl = {https://ui.adsabs.harvard.edu/abs/1974ApJ...187..425P}
}

@ARTICLE{Leonard1990,
       author = {{Leonard}, Peter J.~T. and {Tremaine}, Scott},
        title = "{The Local Galactic Escape Speed}",
      journal = {\apj},
         year = 1990,
        month = apr,
       volume = {353},
        pages = {486},
          doi = {10.1086/168638},
       adsurl = {https://ui.adsabs.harvard.edu/abs/1990ApJ...353..486L}
}
\bibliographystyle{aasjournal}

\end{document}